\documentclass[preprint,prd,aps,showpacs,showkeys,nofootinbib]{revtex4}
\usepackage{cancel}
\usepackage{ulem}
\usepackage{color}
\usepackage{CJK}
\usepackage{graphicx}
\usepackage{dcolumn}
\usepackage{bm}
\begin{document}
\begin{CJK*}{GB}{gbsn}

\title{Corrections to $R_{D}$ and $R_{D^{*}}$ in the BLMSSM}

\author{Zhong-Jun Yang(ÑîÖÒ¾ü)$^{1}$\footnote{zj\_yang1993@163.com}, Shu-Min Zhao(ÕÔÊ÷Ãñ)$^{1}$\footnote{zhaosm@hbu.edu.cn},\\ Xing-Xing Dong(¶­ÐÒÐÒ)$^{1}$\footnote{dxx\_0304@163.com}, Xi-Jie Zhan(չϣ½Ü)$^{1}$,\\ Hai-Bin Zhang(Õź£±ó)$^{1}$, Tai-Fu Feng(·ëÌ«¸µ)$^{1}$}

\affiliation{$^1$ National-Local Joint Engineering Laboratory of New Energy Photoelectric Devices, Department of Physics, Hebei University, Baoding 071002,
China}
\date{\today}
\begin{abstract}
The deviation of the measurement of $R_{D}$ ($R_{D^{*}}$) from the Standard Model (SM) expectation is $2.3\sigma$ ($3.1\sigma$). $R_{D}$ ($R_{D^{*}}$) is the ratio of the branching fraction of $\overline{B} \rightarrow D\tau\overline{\nu}_{\tau}$ ($\overline{B} \rightarrow D^{*}\tau\overline{\nu}_{\tau}$) to that of $\overline{B} \rightarrow Dl\overline{\nu}_{l}$ ($\overline{B} \rightarrow D^{*}l\overline{\nu}_{l}$), where $l = e$ or $\mu$. This anomaly may imply the existence of new physics (NP). In this paper, we restudy this problem in the supersymmetric extension of the Standard Model with local gauged baryon and lepton numbers (BLMSSM), and give one-loop corrections to $R_{D}$ ($R_{D^{*}}$).
\end{abstract}
\maketitle
\newpage
\tableofcontents
\newpage
\end{CJK*}
\section{Introduction}
The Standard Model (SM) is the most successful particle physics model to date. It gives accurate predictions for a significant number of experiments. However, for some experiments, it cannot give a good explanation.
In the last few years, the experimental measurements of $R_{D^{(*)}}$ ( the ratio of the branching fraction of $\overline{B} \rightarrow D\tau\overline{\nu}_{\tau}$ ($\overline{B} \rightarrow D^{*}\tau\overline{\nu}_{\tau}$) to that of $\overline{B} \rightarrow Dl\overline{\nu}_{l}$ ($\overline{B} \rightarrow D^{*}l\overline{\nu}_{l}$), where $l = e$ or $\mu$ ) show deviations from the SM theoretical predictions - these measurements are larger than SM expectations. Therefore, in order to explain these anomalies, it is necessary for us to try some new physics (NP) models.

The SM expectations for $R_{D^{(*)}}$ are: ${R_{D}}_{_\mathcal{SM}}=0.299\pm0.011$ in Ref. \cite{1}, ${R_{D}}_{_\mathcal{SM}}=0.299\pm0.003$ in Ref. \cite{2}, ${R_{D}}_{_\mathcal{SM}}=0.300\pm0.008$ in Ref. \cite{3}, ${R_{D}}_{_\mathcal{SM}}=0.300\pm0.011$ in Ref. \cite{A}, ${R_{D}}_{_\mathcal{SM}}=0.299\pm0.003$ in Ref. \cite{tianjia1}, ${R_{D^{*}}}_{_\mathcal{SM}}=0.254\pm0.004$ in Ref. \cite{A}, ${R_{D^{*}}}_{_\mathcal{SM}}=0.257\pm0.003$ in Ref. \cite{tianjia1} and ${R_{D^{*}}}_{_\mathcal{SM}}=0.252\pm0.003$ in Ref. \cite{4}. The relevant experimental results for $R_{D^{(*)}}$ are listed in the TABLE \ref{tab1}.
\begin{table}[!h]
\caption{ \label{tab1}  The measurements of $R_{D^{(*)}}$.}
\begin{tabular}{|c|c|c|}
  \hline
  ~~~~~~~~~Observable~~~~~~~~~ & ~~~~~~~~~Experiment~~~~~~~~~ & ~~~~~~~~~Measured value~~~~~~~~~ \\
  \hline
  ~ & 2012~BaBar & $0.440\pm 0.058\pm 0.042$ \cite{5,6}  \\
  $R_{D}$ & 2015~Belle & $0.375\pm 0.064\pm 0.026$ \cite{7}  \\
  ~ & 2017~HFAG average & $0.407\pm 0.039\pm 0.024$ \cite{12} \\
  \hline
  ~ & 2012~BaBar & $0.332\pm 0.024\pm 0.018$ \cite{5,6}  \\
  ~ & 2015~Belle & $0.293\pm 0.038\pm 0.015$ \cite{7} \\
  ~ & 2015~LHCb & $0.336\pm 0.027\pm 0.030$ \cite{9} \\
  $R_{D^{*}}$ & 2016~Belle & $0.302\pm 0.030\pm 0.011$ \cite{8}  \\
  ~ & 2017~Belle & $0.270\pm 0.035_{-0.025}^{+0.028}$ \cite{10}  \\
  ~ & 2017~LHCb & $0.291\pm 0.019\pm 0.026\pm 0.013$ \cite{11} \\
  ~ & 2017~HFAG average & $0.304\pm 0.013\pm 0.007$ \cite{12} \\
  \hline
\end{tabular}
\end{table}

$R_{D}=0.407\pm 0.039\pm 0.024$ and $R_{D^{*}}=0.304\pm 0.013\pm 0.007$ exceed the SM predictions by $2.3\sigma$ and $3.1\sigma$ respectively. These anomalies have caused physicists to seek a variety of ways to explain the experimental data \cite{tianjia2,30,31,32,33,34,35,37,38,39,40,41,42,13,14,15,16,17,18,19,20}. Most physicists tend to seek the solutions in NP models. So, various NP models have been used, such as charged Higgs \cite{15,16,17} and lepton flavor violation \cite{18,19,20}. The supersymmetric extension of the SM is a popular choice in various NP models. In fact, theorists have been fond of the minimal supersymmetric model (MSSM) for a long time. However, baryon number (B) should be broken because of the matter-antimatter asymmetry in the Universe. The neutrino oscillation experiments imply that neutrinos have tiny masses, therefore lepton number (L) also needs to be broken. A minimal supersymmetric extension of the SM with local gauged B and L (BLMSSM) \cite{BL41,BL42} is more promising. Thus, we try to deal with the anomalies of $R_{D^{(*)}}$ in the BLMSSM.

In our work, we use effective field theory to do the theoretical calculation. The effective Lagrangian is described by the four fermion operators and the corresponding Wilson coefficients (WCs). NP contributions with non-zero WCs are possible solutions to the $R_{D^{(*)}}$ anomalies \cite{xj1}. After considering all the 10 independent 6-dimensional operators and calculating the values of the corresponding WCs at one-loop level, we obtain the theoretical values of $R_{D^{(*)}}$ in the BLMSSM.

This paper is organised as follows. In section II, we introduce some content of the BLMSSM. In section III, we give the mass matrices of the BLMSSM particles that we use. In section IV, we write down the needed couplings. In section V, we provide the relevant formulae, including observables $R_{D^{(*)}}$ and the effective Lagrangian with all the four fermion operators. In section VI, we show the one-loop Feynman diagrams that can correct $R_{D^{(*)}}$. At the same time, NP contributions of some diagrams are given by WCs. In section VII, we present our numerical results. Finally, we summarise our findings in section VIII. Some integral formulae are shown in the Appendix.
\section{Some content of the BLMSSM}
As an extension of the MSSM, the BLMSSM includes many new fields \cite{BL1,BL2}. The exotic quarks ($\hat{Q}_{4},\hat{U}_{4}^c,\hat{D}_{4}^c,\hat{Q}_{5}^c,\hat{U}_{5},\hat{D}_{5}$) are used to deal with the B anomaly. The exotic leptons ($\hat{L}_{4},\hat{E}_{4}^c,\hat{N}_{4}^c,\hat{L}_{5}^c,\hat{E}_{5},\hat{N}_{5}$) are used to cancel the L anomaly. The exotic Higgs superfields $\hat{\Phi}_{B},\hat{\varphi}_{B}$ are introduced to break baryon number spontaneously with nonzero vacuum expectation values (VEVs). The exotic Higgs superfields $\hat{\Phi}_{L},\hat{\varphi}_{L}$ are introduced to break lepton number spontaneously with non-zero VEVs. The model introduces the right-handed neutrinos $N^{c}_{R}$, so we can obtain tiny masses of neutrinos through the see-saw mechanism. The model also includes the superfields $\hat{X}$ to make the exotic quarks unstable.

 The superpotential of the BLMSSM is \cite{BL3}:
\begin{eqnarray}
&&{\cal W}_{{BLMSSM}}={\cal W}_{{MSSM}}+{\cal W}_{B}+{\cal W}_{L}+{\cal W}_{X}\;,
\label{superpotential1}
\nonumber\\&&{\cal W}_{B}=\lambda_{Q}\hat{Q}_{4}\hat{Q}_{5}^c\hat{\Phi}_{B}+\lambda_{U}\hat{U}_{4}^c\hat{U}_{5}
\hat{\varphi}_{B}+\lambda_{D}\hat{D}_{4}^c\hat{D}_{5}\hat{\varphi}_{B}+\mu_{B}\hat{\Phi}_{B}\hat{\varphi}_{B}
\nonumber\\
&&\hspace{1.2cm}
+Y_{{u_4}}\hat{Q}_{4}\hat{H}_{u}\hat{U}_{4}^c+Y_{{d_4}}\hat{Q}_{4}\hat{H}_{d}\hat{D}_{4}^c
+Y_{{u_5}}\hat{Q}_{5}^c\hat{H}_{d}\hat{U}_{5}+Y_{{d_5}}\hat{Q}_{5}^c\hat{H}_{u}\hat{D}_{5}\;,
\nonumber\\
&&{\cal W}_{L}=Y_{{e_4}}\hat{L}_{4}\hat{H}_{d}\hat{E}_{4}^c+Y_{{\nu_4}}\hat{L}_{4}\hat{H}_{u}\hat{N}_{4}^c
+Y_{{e_5}}\hat{L}_{5}^c\hat{H}_{u}\hat{E}_{5}+Y_{{\nu_5}}\hat{L}_{5}^c\hat{H}_{d}\hat{N}_{5}
\nonumber\\
&&\hspace{1.2cm}
+Y_{\nu}\hat{L}\hat{H}_{u}\hat{N}^c+\lambda_{{N^c}}\hat{N}^c\hat{N}^c\hat{\varphi}_{L}
+\mu_{L}\hat{\Phi}_{L}\hat{\varphi}_{L}\;,
\nonumber\\
&&{\cal W}_{X}=\lambda_1\hat{Q}\hat{Q}_{5}^c\hat{X}+\lambda_2\hat{U}^c\hat{U}_{5}\hat{X}^\prime
+\lambda_3\hat{D}^c\hat{D}_{5}\hat{X}^\prime+\mu_{X}\hat{X}\hat{X}^\prime\;,
\label{superpotential-BL}
\end{eqnarray}
where ${\cal W}_{{MSSM}}$ is the superpotential of the MSSM.

The soft breaking terms $\mathcal{L}_{{soft}}$ of the BLMSSM can be written in the following form \cite{BL3,BL41,BL42} :
\begin{eqnarray}
&&{\cal L}_{{soft}}={\cal L}_{{soft}}^{MSSM}-(m_{{\tilde{\nu}^c}}^2)_{{IJ}}\tilde{N}_I^{c*}\tilde{N}_J^c
-m_{{\tilde{Q}_4}}^2\tilde{Q}_{4}^\dagger\tilde{Q}_{4}-m_{{\tilde{U}_4}}^2\tilde{U}_{4}^{c*}\tilde{U}_{4}^c
-m_{{\tilde{D}_4}}^2\tilde{D}_{4}^{c*}\tilde{D}_{4}^c
\nonumber\\
&&\hspace{1.3cm}
-m_{{\tilde{Q}_5}}^2\tilde{Q}_{5}^{c\dagger}\tilde{Q}_{5}^c-m_{{\tilde{U}_5}}^2\tilde{U}_{5}^*\tilde{U}_{5}
-m_{{\tilde{D}_5}}^2\tilde{D}_{5}^*\tilde{D}_{5}-m_{{\tilde{L}_4}}^2\tilde{L}_{4}^\dagger\tilde{L}_{4}
-m_{{\tilde{\nu}_4}}^2\tilde{N}_{4}^{c*}\tilde{N}_{4}^c
\nonumber\\
&&\hspace{1.3cm}
-m_{{\tilde{e}_4}}^2\tilde{E}_{_4}^{c*}\tilde{E}_{4}^c-m_{{\tilde{L}_5}}^2\tilde{L}_{5}^{c\dagger}\tilde{L}_{5}^c
-m_{{\tilde{\nu}_5}}^2\tilde{N}_{5}^*\tilde{N}_{5}-m_{{\tilde{e}_5}}^2\tilde{E}_{5}^*\tilde{E}_{5}
-m_{{\Phi_{B}}}^2\Phi_{B}^*\Phi_{B}
\nonumber\\
&&\hspace{1.3cm}
-m_{{\varphi_{B}}}^2\varphi_{B}^*\varphi_{B}-m_{{\Phi_{L}}}^2\Phi_{L}^*\Phi_{L}
-m_{{\varphi_{L}}}^2\varphi_{L}^*\varphi_{L}-\Big(M_{B}\lambda_{B}\lambda_{B}
+M_{L}\lambda_{L}\lambda_{L}+h.c.\Big)
\nonumber\\
&&\hspace{1.3cm}
+\Big\{A_{{u_4}}Y_{{u_4}}\tilde{Q}_{4}H_{u}\tilde{U}_{4}^c+A_{{d_4}}Y_{{d_4}}\tilde{Q}_{4}H_{d}\tilde{D}_{4}^c
+A_{{u_5}}Y_{{u_5}}\tilde{Q}_{5}^cH_{d}\tilde{U}_{5}+A_{{d_5}}Y_{{d_5}}\tilde{Q}_{5}^cH_{u}\tilde{D}_{5}
\nonumber\\
&&\hspace{1.3cm}
+A_{{BQ}}\lambda_{Q}\tilde{Q}_{4}\tilde{Q}_{5}^c\Phi_{B}+A_{{BU}}\lambda_{U}\tilde{U}_{4}^c\tilde{U}_{5}\varphi_{B}
+A_{{BD}}\lambda_{D}\tilde{D}_{4}^c\tilde{D}_{5}\varphi_{B}+B_{B}\mu_{B}\Phi_{B}\varphi_{B}
+h.c.\Big\}
\nonumber\\
&&\hspace{1.3cm}
+\Big\{A_{{e_4}}Y_{{e_4}}\tilde{L}_{4}H_{d}\tilde{E}_{4}^c+A_{{\nu_4}}Y_{{\nu_4}}\tilde{L}_{4}H_{u}\tilde{N}_{4}^c
+A_{{e_5}}Y_{{e_5}}\tilde{L}_{5}^cH_{u}\tilde{E}_{5}+A_{{\nu_5}}Y_{{\nu_5}}\tilde{L}_{5}^cH_{d}\tilde{N}_{5}
\nonumber\\
&&\hspace{1.3cm}
+A_{\nu}Y_{\nu}\tilde{L}H_{u}\tilde{N}^c+A_{{\nu^c}}\lambda_{{\nu^c}}\tilde{N}^c\tilde{N}^c\varphi_{L}
+B_{L}\mu_{L}\Phi_{L}\varphi_{L}+h.c.\Big\}
\nonumber\\
&&\hspace{1.3cm}
+\Big\{A_1\lambda_1\tilde{Q}\tilde{Q}_{5}^cX+A_2\lambda_2\tilde{U}^c\tilde{U}_{5}X^\prime
+A_3\lambda_3\tilde{D}^c\tilde{D}_{5}X^\prime+B_{X}\mu_{X}XX^\prime+h.c.\Big\}\;.
\label{soft-breaking}
\end{eqnarray}

The $SU(2)_L$ singlets $\Phi_{L},\;\varphi_{L},\;\Phi_{B},\;\varphi_{B}$ and the $SU(2)_{L}$ doublets $H_{u},\;H_{d}$ are:
\begin{eqnarray}
&&\Phi_{L}={1\over\sqrt{2}}\Big(\upsilon_{L}+\Phi_{L}^0+iP_{L}^0\Big)\;,~~~~~~~~~
\varphi_{L}={1\over\sqrt{2}}\Big(\overline{\upsilon}_{L}+\varphi_{L}^0+i\overline{P}_{L}^0\Big)\;,
\nonumber\\
&&\Phi_{B}={1\over\sqrt{2}}\Big(\upsilon_{B}+\Phi_{B}^0+iP_{B}^0\Big)\;,~~~~~~~~
\varphi_{B}={1\over\sqrt{2}}\Big(\overline{\upsilon}_{B}+\varphi_{B}^0+i\overline{P}_{B}^0\Big)\;,\nonumber
\end{eqnarray}
\begin{eqnarray}
&&H_{u}=\left(\begin{array}{c}H_{u}^+\\{1\over\sqrt{2}}\Big(\upsilon_{u}+H_{u}^0+iP_{u}^0\Big)\end{array}\right)\;,~~~~
H_{d}=\left(\begin{array}{c}{1\over\sqrt{2}}\Big(\upsilon_{d}+H_{d}^0+iP_{d}^0\Big)\\H_{d}^-\end{array}\right)\;.
\label{VEVs}
\end{eqnarray}

The $SU(2)_L$ singlets $\Phi_{L},\;\varphi_{L},\;\Phi_{B},\;\varphi_{B}$ and the $SU(2)_{L}$ doublets $H_{u},\;H_{d}$ should obtain non-zero VEVs
 $\upsilon_{L},\;\overline{\upsilon}_{L},\;\upsilon_{{B}},\;\overline{\upsilon}_{{B}}$
and $\upsilon_{u},\;\upsilon_{d}$ respectively. Therefore, the local gauge symmetry $SU(2)_{L}\otimes U(1)_{Y}\otimes U(1)_{B}\otimes U(1)_{L}$
breaks down to the electromagnetic symmetry $U(1)_{e}$.
\section{Mass matrices for some BLMSSM particles}
Lepneutralinos are made up of $\lambda_L$ (the superpartner of the new lepton boson), and $\psi_{\Phi_L}$ and $\psi_{\varphi_L}$
 (the superpartners of the $SU(2)_L$ singlets $\Phi_L$ and $\varphi_L$). The mass mixing matrix of lepneutralinos $M_{LN}$ is shown in the
 basis $(i\lambda_L,\psi_{\Phi_L},\psi_{\varphi_L})$ \cite{BL5,BL6,BL7,BL8}. $\chi^0_{L_i}~(i=1,2,3)$ are mass eigenstates of lepneutralinos. The masses of the three lepneutralinos are obtained from diagonalizing $M_{LN}$ by $Z_{N_L}$:
 \begin{eqnarray}
&& M_{LN}=\left( \begin{array}{ccc}
  2M_L &2v_Lg_L &-2\bar{v}_Lg_L\\
   2v_Lg_L & 0 &-\mu_L\\-2\bar{v}_Lg_L&-\mu_L &0
    \end{array}\right),
    \nonumber\\&& i\lambda_L=Z_{N_L}^{1i}k_{L_i}^0,~~
   \psi_{\Phi_L}=Z_{N_L}^{2i}k_{L_i}^0,~~\nonumber\\&&
   \psi_{\varphi_L}=Z_{N_L}^{3i}k_{L_i}^0,~~ \chi^0_{L_i}= \left(\begin{array}{c}
 k_{L_i}^0\\ \bar{k}_{L_i}^0
    \end{array}\right).\label{neutrinoD}
  \end{eqnarray}

The slepton mass squared matrix becomes
\begin{eqnarray}
&&\left(\begin{array}{cc}
  (\mathcal{M}^2_{\tilde{L}})_{LL}&(\mathcal{M}^2_{\tilde{L}})_{LR} \\
   (\mathcal{M}^2_{\tilde{L}})_{LR}^{\dag} & (\mathcal{M}^2_{\tilde{L}})_{RR}
    \end{array}\right),
\end{eqnarray}
which is diagonalized by the matrix $Z_{\tilde{L}}$. $(\mathcal{M}^2_{\tilde{L}})_{LL},~(\mathcal{M}^2_{\tilde{L}})_{LR}$ and $(\mathcal{M}^2_{\tilde{L}})_{RR}$ are:
\begin{eqnarray}
 &&(\mathcal{M}^2_{\tilde{L}})_{LL}=\frac{(g_1^2-g_2^2)(v_d^2-v_u^2)}{8}\delta_{IJ} +g_L^2(\bar{v}_L^2-v_L^2)\delta_{IJ}
 +m_{l^I}^2\delta_{IJ}+(m^2_{\tilde{L}})_{IJ},\nonumber\\&&
 (\mathcal{M}^2_{\tilde{L}})_{LR}=\frac{\mu^*v_u}{\sqrt{2}}(Y_l)_{IJ}-\frac{v_u}{\sqrt{2}}(A'_l)_{IJ}+\frac{v_d}{\sqrt{2}}(A_l)_{IJ},
 \nonumber\\&& (\mathcal{M}^2_{\tilde{L}})_{RR}=\frac{g_1^2(v_u^2-v_d^2)}{4}\delta_{IJ}-g_L^2(\bar{v}_L^2-v_L^2)\delta_{IJ}
 +m_{l^I}^2\delta_{IJ}+(m^2_{\tilde{R}})_{IJ}.
\end{eqnarray}

The mass squared matrix of sneutrino ${\cal M}_{\tilde{n}}$ with $\tilde{n}^T=(\tilde{\nu},\tilde{N}^c)$ reads \cite{xj2}
\begin{eqnarray}
&&\left(\begin{array}{cc}
  {\cal M}^2_{\tilde{n}}(\tilde{\nu}_{I}^*\tilde{\nu}_{J})&{\cal M}^2_{\tilde{n}}(\tilde{\nu}_I\tilde{N}_J^c) \\
   ({\cal M}^2_{\tilde{n}}(\tilde{\nu}_I\tilde{N}_J^c))^{\dag} & {\cal M}^2_{\tilde{n}}(\tilde{N}_I^{c*}\tilde{N}_J^c)
    \end{array}\right).
\end{eqnarray}
${\cal M}^2_{\tilde{n}}(\tilde{\nu}_{I}^*\tilde{\nu}_{J})$, ${\cal M}^2_{\tilde{n}}(\tilde{\nu}_I\tilde{N}_J^c)$ and ${\cal M}^2_{\tilde{n}}(\tilde{N}_I^{c*}\tilde{N}_J^c)$ are:
\begin{eqnarray}
  && {\cal M}^2_{\tilde{n}}(\tilde{\nu}_{I}^*\tilde{\nu}_{J})=\frac{g_1^2+g_2^2}{8}(v_d^2-v_u^2)\delta_{IJ}+g_L^2(\overline{v}^2_L-v^2_L)\delta_{IJ}
   +\frac{v_u^2}{2}(Y^\dag_{\nu}Y_\nu)_{IJ}+(m^2_{\tilde{L}})_{IJ},\nonumber\\&&
   {\cal M}^2_{\tilde{n}}(\tilde{\nu}_I\tilde{N}_J^c)=\mu^*\frac{v_d}{\sqrt{2}}(Y_{\nu})_{IJ}-v_u\overline{v}_L(Y_{\nu}^\dag\lambda_{N^c})_{IJ}
   +\frac{v_u}{\sqrt{2}}(A_{N})_{IJ}(Y_\nu)_{IJ},\nonumber\\&&
   {\cal M}^2_{\tilde{n}}(\tilde{N}_I^{c*}\tilde{N}_J^c)=-g_L^2(\overline{v}^2_L-v^2_L)\delta_{IJ}
   +\frac{v_u^2}{2}(Y^\dag_{\nu}Y_\nu)_{IJ}+2\overline{v}^2_L(\lambda_{N^c}^\dag\lambda_{N^c})_{IJ}\nonumber\\&&
   \hspace{2.9cm}+(m^2_{\tilde{N}^c})_{IJ}+\mu_L\frac{v_L}{\sqrt{2}}(\lambda_{N^c})_{IJ}
   -\frac{\overline{v}_L}{\sqrt{2}}(A_{N^c})_{IJ}(\lambda_{N^c})_{IJ}.
   \end{eqnarray}
Then the masses of the sneutrinos are obtained by using the formula
$Z_{\tilde{\nu}}^{\dag} {\cal M}_{\tilde{n}}^2Z_{\tilde{\nu}}=diag(m_{\tilde{\nu}^1}^2, m_{\tilde{\nu}^2}^2, m_{\tilde{\nu}^3}^2, m_{\tilde{\nu}^4}^2, m_{\tilde{\nu}^5}^2, m_{\tilde{\nu}^6}^2)$.

The up scalar quark mass squared matrix in the BLMSSM is given by
\begin{eqnarray}
&&\left(\begin{array}{cc}
  (\mathcal{M}^2_{\tilde{U}})_{LL}&(\mathcal{M}^2_{\tilde{U}})_{LR} \\
   (\mathcal{M}^2_{\tilde{U}})_{LR}^{\dag} & (\mathcal{M}^2_{\tilde{U}})_{RR}
    \end{array}\right),
\end{eqnarray}
which is diagonalized by the matrix $Z_{\tilde{U}}$. $(\mathcal{M}^2_{\tilde{U}})_{LL},~(\mathcal{M}^2_{\tilde{U}})_{LR}$ and $(\mathcal{M}^2_{\tilde{U}})_{RR}$ are:
\begin{eqnarray}
&&(M^2_{\tilde{U}})_{LL}=-\frac{e^2(v_d^2-v_u^2)(1-4c_W^2)}{24s_W^2c_W^2}+\frac{v_u^2Y_u^2}{2} +(Km_{\tilde{Q}}^2K^\dag)^T+\frac{g_B^2}{6}(v_B^2-\bar{v}^2_B),
\nonumber\\&&(M^2_{\tilde{U}})_{RR}=\frac{e^2(v_d^2-v_u^2)}{6c_W^2}+\frac{v_u^2Y_u^2}{2} +m_{\tilde{U}}^2-\frac{g_B^2}{6}(v_B^2-\bar{v}^2_B),
\nonumber\\&&(M^2_{\tilde{U}})_{LR}=-\frac{1}{\sqrt{2}}\Big(v_d(A_u'+Y_u\mu^*)+v_uA_u\Big).
\end{eqnarray}

The down scalar quark mass squared matrix in the BLMSSM is given by
\begin{eqnarray}
&&\left(\begin{array}{cc}
  (\mathcal{M}^2_{\tilde{D}})_{LL}&(\mathcal{M}^2_{\tilde{D}})_{LR} \\
   (\mathcal{M}^2_{\tilde{D}})_{LR}^{\dag} & (\mathcal{M}^2_{\tilde{D}})_{RR}
    \end{array}\right),
\end{eqnarray}
which is diagonalized by the matrix $Z_{\tilde{D}}$. $(\mathcal{M}^2_{\tilde{U}})_{LL},~(\mathcal{M}^2_{\tilde{U}})_{LR}$ and $(\mathcal{M}^2_{\tilde{U}})_{RR}$ are:
\begin{eqnarray}
&&(M^2_{\tilde{D}})_{LL}=-\frac{e^2(v_d^2-v_u^2)(1+2c_W^2)}{24s_W^2c_W^2}+\frac{v_d^2Y_d^2}{2} +(m_{\tilde{Q}}^2)^T+\frac{g_B^2}{6}(v_B^2-\bar{v}^2_B),
\nonumber\\&&(M^2_{\tilde{D}})_{RR}=-\frac{e^2(v_d^2-v_u^2)}{12c_W^2}+\frac{v_d^2Y_d^2}{2} +m_{\tilde{D}}^2-\frac{g_B^2}{6}(v_B^2-\bar{v}^2_B),
\nonumber\\&&(M^2_{\tilde{D}})_{LR}=\frac{1}{\sqrt{2}}\Big(v_u(-A_d'+Y_d\mu^*)+v_dA_d\Big).
\end{eqnarray}

In the basis $( \psi_ {\nu^I_L},\psi_{N^{cI}_R} )$, the neutrino mass mixing matrix is diagonalized by $Z_{\nu}$ \cite{xj2}:
\begin{eqnarray}
&&Z_{\nu}^{T}\left(\begin{array}{cc}
  0&\frac{v_u}{\sqrt{2}}(Y_{\nu})^{IJ} \\
   \frac{v_u}{\sqrt{2}}(Y^{T}_{\nu})^{IJ}  & \frac{\bar{v}_L}{\sqrt{2}}(\lambda_{N^c})^{IJ}
    \end{array}\right) Z_{\nu}
  \nonumber\\&&=diag(m_{\nu^\alpha}), ~~~~ \alpha=1\dots 6.
    \nonumber\\&& \psi_{\nu^I_L}=Z_{\nu}^{I\alpha}k_{N_\alpha}^0,~~~~
   \psi_{N^{cI}_R}=Z_{\nu}^{(I+3)\alpha}k_{N_\alpha}^0,~~~~\nonumber\\
   &&\nu^\alpha= \left(\begin{array}{c}
 k_{N_\alpha}^0\\ \bar{k}_{N_\alpha}^0
    \end{array}\right).\label{neutrinoD}
      \end{eqnarray}
 $\nu^\alpha$ denotes the mass eigenstates of the neutrino fields mixed by the left-handed and right-handed neutrinos. In this paper, we deal with the neutrinos by an approximation, $Z_{\nu} \approx 1$, so the theoretical values at tree level are consistent with those in the SM.
\section{Necessary couplings}
In the BLMSSM, due to the superfields $\tilde{N}^c$, we deduce the corrections to the couplings in the MSSM. The couplings for $W$-$l$-$\nu$ and $W$-$\tilde{L}$-$\tilde{\nu}$ read
\begin{eqnarray}
&&\mathcal{L}_{Wl\nu}=-\frac{e}{\sqrt{2}s_W}W_{\mu}^+\sum_{I=1}^3\sum_{\alpha=1}^6 Z_{\nu}^{I \alpha *}\bar{\nu}^\alpha \gamma^{\mu}P_Ll^I,\\&&\hspace{0cm}
\mathcal{L}_{W\tilde{L}\tilde{\nu}}=-\frac{ie}{\sqrt{2}s_W}W_{\mu}^-\sum_{I=1}^3\sum_{i,\alpha=1}^6 (Z_{\tilde{L}}^{I i} Z_{\tilde{\nu}}^{I \alpha})(\tilde{L}_{i}^{+} (\overrightarrow{\partial^{\mu}}-\overleftarrow{\partial^{\mu}})\tilde{\nu}^{\alpha}).
\end{eqnarray}

From the interactions of gauge and matter multiplets
$ig\sqrt{2}T^a_{ij}(\lambda^a\psi_jA_i^*-\bar{\lambda}^a\bar{\psi}_iA_j)$,
the $l$-$\chi_L^0$-$\tilde{L}$ coupling is deduced here:
\begin{eqnarray}
&&\mathcal{L}_{l\chi_L^0\tilde{L}}=\sqrt{2}g_L\bar{\chi}_{L_j}^0\Big(Z_{N_L}^{1j}Z_{\tilde{L}}^{Ii}P_L
-Z_{N_L}^{1j*}Z_{\tilde{L}}^{(I+3)i}P_R\Big)l^I\tilde{L}_i^++h.c.
\end{eqnarray}
The $\nu$-$\chi_L^0$-$\tilde{\nu}$ coupling is
\begin{eqnarray}
&&\mathcal{L}_{\nu\chi_L^0\tilde{\nu}}=[\sqrt{2}g_LZ_{N_L}^{1i}Z_{\nu}^{I\alpha}Z_{\tilde{\nu}}^{Jj*}\delta^{IJ}
-(Z_{N_L}^{3i}(\lambda^{IJ}_{N^{c}}+\lambda^{JI}_{N^{c}})+\sqrt{2}g_LZ_{N_L}^{1i}\delta^{IJ})\nonumber\\&&\hspace{1.7cm}\times Z_{\nu}^{(I+3)\alpha}Z_{\tilde{\nu}}^{(J+3)j*}]\bar{\chi}_{L_i}^0P_L\nu^{\alpha}\tilde{\nu}^{j*}+h.c.
\end{eqnarray}

We also obtain the $\chi^{\pm}$-$l$-$\tilde{\nu}$ coupling and the $\chi^{\pm}$-$\tilde{L}$-$\nu$ coupling:
\begin{eqnarray}
&&\mathcal{L}_{\chi^{\pm}l\tilde{\nu}}=-\sum_{I,J=1}^3\sum_{\alpha=1}^6\bar{\chi}^-_j
\Big(Y_l^{IJ} Z_-^{2j*}(Z_{\tilde{\nu}}^{I \alpha})^{*}P_R\nonumber\\&&\hspace{1.7cm}
+[\frac{e}{s_W}Z_+^{1j}(Z_{\tilde{\nu}}^{I \alpha})^{*}+Y_\nu^{IJ}Z_+^{2j}(Z_{\tilde{\nu}}^{(I+3) \alpha})^{*}]P_L
\Big)l^J\tilde{\nu}^{\alpha*}+h.c.\nonumber\\&&
\mathcal{L}_{\chi^{\pm}\tilde{L}\nu}=-\sum_{I,J=1}^3\sum_{i=1}^2\sum_{j,\alpha=1}^6\bar{\chi}^+_i
\Big(Y_\nu^{IJ} Z_+^{2i*}Z_{\tilde{L}}^{I j}Z_{\nu}^{(J+3) \alpha *}P_R\nonumber\\&&\hspace{1.7cm}
+[\frac{e}{s_W}Z_-^{1i}Z_{\tilde{L}}^{I j}+Y_l^{IJ}Z_-^{2i}Z_{\tilde{L}}^{(I+3) j}] Z_{\nu}^{J \alpha}P_L
\Big)\nu^{\alpha}\tilde{L}^{+}_j+h.c.
\end{eqnarray}

The $\chi^0$-$\tilde{\nu}$-$\nu$ coupling in the BLMSSM becomes
\begin{eqnarray}
&&\mathcal{L}_{\chi^0\tilde{\nu}\nu}=[Z_{\nu}^{I \alpha}Z_{\tilde{\nu}}^{J j*}\frac{e}{\sqrt{2} s_W c_W}(Z_{N}^{1i}s_W-Z_{N}^{2i}c_W)\nonumber\\&&\hspace{1.7cm}
+\frac{Y_{\nu}^{IJ}}{\sqrt{2}}Z_{N}^{4i}(Z_{\nu}^{I\alpha}Z_{\tilde{\nu}}^{(J+3)j*}+Z_{\nu}^{(I+3)\alpha}Z_{\tilde{\nu}}^{Jj*})]\bar{\chi}_{i}^0P_L\nu^{\alpha}\tilde{\nu}^{j*}+h.c.
\end{eqnarray}

All the other couplings used are consistent with the MSSM.
\section{Formulae}
\subsection{Observables}
The observable $R_{D^{(*)}}$ is defined as
\begin{eqnarray}
&&R_{D^{(*)}}=\frac{\mathcal{B}_{\tau}^{D^{(*)}}}{\mathcal{B}_{l}^{D^{(*)}}}=\frac{\mathcal{B}(\overline{B}\rightarrow D^{(*)}\tau \bar{\nu}_{\tau})}{\mathcal{B}(\overline{B}\rightarrow D^{(*)}l \bar{\nu}_{l})}.
\label{DFRDX}
\end{eqnarray}
$\mathcal{B}_{\ell}^{D^{(*)}}$, the branching fraction, is given by~\cite{A}
\begin{eqnarray}
&&\mathcal{B}_{\ell}^{D^{(*)}}=\int \mathcal{N}|p_{D^{(*)}}|(2a_{\ell}^{D^{(*)}}+\frac{2}{3}c_{\ell}^{D^{(*)}})dq^{2},
\end{eqnarray}
where $l=e$ or $\mu$, and $\ell$ denotes any lepton $(e,\mu$ or $\tau)$. $q^{2}$ is the invariant mass squared of the lepton-neutrino system, whose integral interval is $[m_{\ell}^{2},(M_{B}-M_{D^{(*)}})^{2}]$. $\mathcal{N}$, the normalisation factor, is given by
\begin{eqnarray}
&&\mathcal{N}=\frac{\tau_{B}G_{F}^{2}|V_{cb}|^{2}q^{2}}{256\pi^{3}M_{B}^{2}}(1-\frac{m_{\ell}^{2}}{q^{2}})^2.
\end{eqnarray}
Here $\tau_{B}$ is the lifetime of the $B-$meson. $G_{F}=\sqrt{2}e^2/8m_W^{2}s_{W}^{2}$ is the Fermi coupling constant. $|p_{D^{(*)}}|$, the absolute value of the $D^{(*)}-$meson momentum, is given by
\begin{eqnarray}
&&|p_{D^{(*)}}|=\frac{\sqrt{(M_{B}^{2})^{2}+(M_{D^{(*)}}^{2})^{2}+(q^{2})^{2}-2(M_{B}^{2}M_{D^{(*)}}^{2}+M_{D^{(*)}}^{2}q^{2}+q^{2}M_{B}^{2})}}{2M_{B}}.
\end{eqnarray}

The expressions for $a_{\ell}^{D}$ and $c_{\ell}^{D}$ are \cite{A}:
\begin{eqnarray}
&&a_{\ell}^{D}=8\{\frac{M_{B}^{2}|p_{D}|^{2}}{q^{2}}(|\mathcal{C}_{VL}^{\ell}|^{2}+|\mathcal{C}_{VR}^{\ell}|^{2})\textbf{F}_{+}^{2}+\frac{(M_{B}^{2}- M_{D}^{2})^{2}}{4(m_{b}-m_{c})^{2}}(|\mathcal{C}_{SL}^{\ell}|^{2}+|\mathcal{C}_{SR}^{\ell}|^{2})\textbf{F}_{0}^{2}\nonumber\\&&\hspace{1cm}
+m_{\ell}[\frac{(M_{B}^{2}- M_{D}^{2})^{2}}{2q^{2}(m_{b}-m_{c})}(\mathcal{R}(\mathcal{C}_{SL}^{\ell}\mathcal{C}_{VL}^{\ell*})+\mathcal{R}(\mathcal{C}_{SR}^{\ell}\mathcal{C}_{VR}^{\ell*}))\textbf{F}_{0}^{2}\nonumber\\&&\hspace{1cm}
+\frac{4M_{B}^{2} |p_{D}|^{2} }{q^{2}(M_{B}+M_{D})}(\mathcal{R}(\mathcal{C}_{TL}^{\ell}\mathcal{C}_{VL}^{\ell*})+\mathcal{R}(\mathcal{C}_{TR}^{\ell}\mathcal{C}_{VR}^{\ell*}))\textbf{F}_{+}\textbf{F}_{T}]\nonumber\\&&\hspace{1cm}
+m_{\ell}^{2}[\frac{(M_{B}^{2}- M_{D}^{2})^{2}}{4q^{4}}(|\mathcal{C}_{VL}^{\ell}|^{2}+|\mathcal{C}_{VR}^{\ell}|^{2})\textbf{F}_{0}^{2}\nonumber\\&&\hspace{1cm}
+\frac{4|p_{D}|^{2}M_{B}^{2}  }{q^{2}(M_{B}+M_{D})^{2}}(| \mathcal{C}_{TL}^{\ell}|^{2}+| \mathcal{C}_{TR}^{\ell}|^{2})\textbf{F}_{T}^{2}]\},
\end{eqnarray}
\begin{eqnarray}
&&c_{\ell}^{D}=8\{\frac{4M_{B}^{2}|p_{D}|^{2}}{(M_{B}+ M_{D})^{2}}(|\mathcal{C}_{TL}^{\ell}|^{2}+|\mathcal{C}_{TR}^{\ell}|^{2})\textbf{F}_{T}^{2}\nonumber\\&&\hspace{1cm}
-\frac{M_{B}^{2} |p_{D}|^{2}}{q^{2}}(|\mathcal{C}_{VL}^{\ell}|^{2}+|\mathcal{C}_{VR}^{\ell}|^{2})\textbf{F}_{+}^{2}
+m_{\ell}^{2}[\frac{|p_{D}|^{2}M_{B}^{2}}{q^{4}}(|\mathcal{C}_{VL}^{\ell}|^{2}+|\mathcal{C}_{VR}^{\ell}|^{2})\textbf{F}_{+}^{2}\nonumber\\&&\hspace{1cm}
-\frac{4|p_{D}|^{2}M_{B}^{2}}{(M_{B}+M_{D})^{2}q^{2}}(|\mathcal{C}_{TL}^{\ell}|^{2}+| \mathcal{C}_{TR}^{\ell}|^{2})\textbf{F}_{T}^{2}]\}.
\end{eqnarray}
The full expressions for $a_{\ell}^{D^{*}}$, $c_{\ell}^{D^{*}}$ and all form factors ($\textbf{F}_{T}(q^{2})$, $\textbf{F}_{+}(q^{2})$ and $\textbf{F}_{0}(q^{2})$, etc) are given in Refs. \cite{A,36}.
\subsection{Effective Lagrangian}
We use effective field theory to calculate the theoretical values. The effective Lagrangian for the $b\rightarrow c\ell\bar{\nu}^{_\ell}$ process is
\begin{eqnarray}
&&\mathcal{L}_{eff}^{b\rightarrow c\ell\bar{\nu}^{_\ell}}=\sqrt{2}G_{F}V_{cb}(\mathcal{C}_{VL}^{\ell}\mathcal{O}_{VL}^{\ell}+\mathcal{C}_{VR}^{\ell}\mathcal{O}_{VR}^{\ell}+\mathcal{C}_{AL}^{\ell}\mathcal{O}_{AL}^{\ell}+\mathcal{C}_{AR}^{\ell}\mathcal{O}_{AR}^{\ell}+\mathcal{C}_{SL}^{\ell}\mathcal{O}_{SL}^{\ell}\nonumber\\&&\hspace{1.7cm}
+\mathcal{C}_{SR}^{\ell}\mathcal{O}_{SR}^{\ell}+\mathcal{C}_{PL}^{\ell}\mathcal{O}_{PL}^{\ell}+\mathcal{C}_{PR}^{\ell}\mathcal{O}_{PR}^{\ell}+\mathcal{C}_{TL}^{\ell}\mathcal{O}_{TL}^{\ell}+\mathcal{C}_{TR}^{\ell}\mathcal{O}_{TR}^{\ell}),
\label{EL}
\end{eqnarray}
where $V_{cb}=0.04$, and the full set of operators is \cite{G}:
\begin{eqnarray}
&&\mathcal{O}_{VL}^{\ell}=[\bar{c}\gamma^\mu b][\bar{\ell}\gamma_\mu P_L\nu^{_\ell}], ~~~~~~~~~~~~
\mathcal{O}_{VR}^{\ell}=[\bar{c}\gamma^\mu b][\bar{\ell}\gamma_\mu P_R\nu^{_\ell}],\nonumber\\&&\hspace{0cm}
\mathcal{O}_{AL}^{\ell}=[\bar{c}\gamma^\mu \gamma_5 b][\bar{\ell}\gamma_\mu P_L\nu^{_\ell}], ~~~~~~~~~~
\mathcal{O}_{AR}^{\ell}=[\bar{c}\gamma^\mu \gamma_5 b][\bar{\ell}\gamma_\mu P_R\nu^{_\ell}],\nonumber\\&&\hspace{0cm}
\mathcal{O}_{SL}^{\ell}=[\bar{c} b][\bar{\ell} P_L\nu^{_\ell}], ~~~~~~~~~~~~~~~~~~~
\mathcal{O}_{SR}^{\ell}=[\bar{c} b][\bar{\ell} P_R\nu^{_\ell}],\nonumber\\&&\hspace{0cm}
\mathcal{O}_{PL}^{\ell}=[\bar{c}\gamma_5 b][\bar{\ell} P_L\nu^{_\ell}], ~~~~~~~~~~~~~~~~
\mathcal{O}_{PR}^{\ell}=[\bar{c}\gamma_5 b][\bar{\ell} P_R\nu^{_\ell}],\nonumber\\&&\hspace{0cm}
\mathcal{O}_{TL}^{\ell}=[\bar{c}\sigma^{\mu\nu} b][\bar{\ell}\sigma_{\mu\nu} P_L\nu^{_\ell}], ~~~~~~~~~~
\mathcal{O}_{TR}^{\ell}=[\bar{c}\sigma^{\mu\nu} b][\bar{\ell}\sigma_{\mu\nu} P_R\nu^{_\ell}].
\end{eqnarray}
In the SM, $\mathcal{C}_{VL}^{\ell}=-\mathcal{C}_{AL}^{\ell}=1$ and all the other WCs vanish. In the BLMSSM, we calculate all the WCs at one-loop level to obtain the theoretical values.
\section{Feynman diagrams }
\begin{figure}[!htbp]
\centering
\includegraphics[width=10cm]{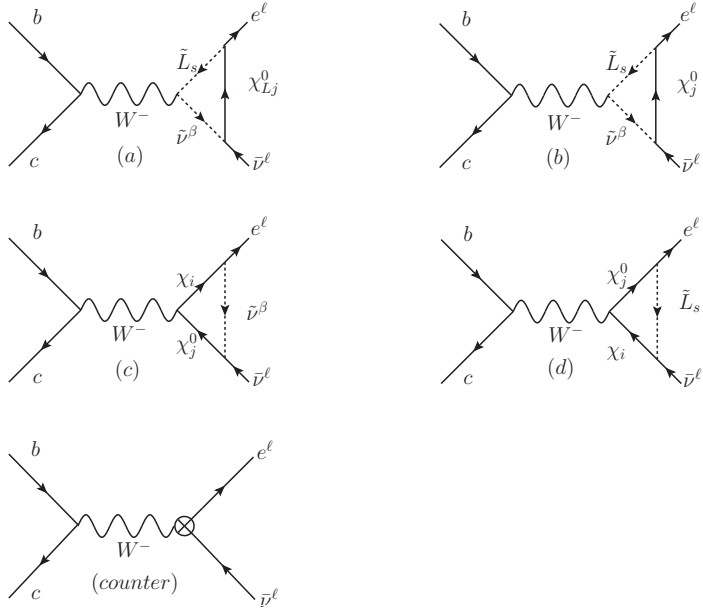}\\
\caption{The penguin-type Feynman diagrams that can correct $R_{D^{(*)}}$ in BLMSSM} \label{fig1}
\end{figure}

In the BLMSSM, the one-loop Feynman diagrams for the lepton sector that can correct the anomalies are shown in FIG.~\ref{fig1} and FIG.~\ref{fig2}.
\subsection{Penguin-type Feynman diagrams}
\subsubsection{The WCs}
The one-loop Feynman diagrams FIG.~\ref{fig1} (a),(b),(c) and (d) are all UV divergent. Focusing on FIG.~\ref{fig1} (a), the three lepneutralinos $\chi^0_{L}$ are new particles in the BLMSSM, and they play very important roles in this decay process. So taking FIG.~\ref{fig1} (a) as an example, the non-zero WCs in Eq. (\ref{EL}) are given as follows:
\begin{eqnarray}
\mathcal{C}_{VL(a)}^{\ell}&=&[\sum_{\beta,s=1}^{6}\sum_{j=1}^{3}\frac{\mathcal{B}_{1}^{\ell sj}\mathcal{A}_{2}^{\beta\ell j}\mathcal{A}_{3}^{\beta s}\mathcal{A}_{4}}{m_{W}^{2}}\frac{1}{64\pi^{2}}(\Delta_{UV}+1\nonumber\\&&\hspace{0cm}
-F_{21}(x_{\chi^0_{L_j}},x_{\tilde{L}_{s}},x_{\tilde{\nu}^{\beta}}))]/(\sqrt{2}G_{F}V_{cb})\nonumber\\&&\hspace{-0.5cm}=\frac{g_L^{2}}{16 \pi^{2}}\Delta_{UV}+finite~terms,\nonumber\\&&\hspace{-1.8cm}
\mathcal{C}_{AL(a)}^{\ell}~=-\mathcal{C}_{VL(a)}^{\ell}.
\label{Ca}
\end{eqnarray}
Here, we use the unitary characteristics of the rotation matrices. In Eq. (\ref{Ca}),
\begin{eqnarray}
&&\mathcal{B}_{1}^{\ell sj}=\sqrt{2}g_{L}Z_{N_{L}}^{1j*}Z_{\tilde{L}}^{\ell s*},\nonumber\\&&\hspace{0cm}
\mathcal{A}_{2}^{\beta\ell j}=\sum_{I=1}^{3}(\sqrt{2}g_{L}Z_{N_{L}}^{1j}Z_{\nu}^{I\ell}Z_{\tilde{\nu}}^{I\beta}-(2\lambda_{N^{c}}^{II}Z_{N_{L}}^{3j}+\sqrt{2}g_{L}Z_{N_{L}}^{1j})Z_{\nu}^{(I+3)\ell}Z_{\tilde{\nu}}^{(I+3)\beta*}),\nonumber\\&&\hspace{0cm}
\mathcal{A}_{3}^{\beta s}=-\sum_{I=1}^{3}(\frac{e}{\sqrt{2}s_{W}}Z_{\tilde{\nu}}^{I\beta}Z_{\tilde{L}}^{Is}),~~~
\mathcal{A}_{4}=-\frac{e}{\sqrt{2}s_{W}}V_{cb}.
\end{eqnarray}
$\Delta_{UV}=1/\epsilon+\ln(4\pi \kappa^2/\Lambda^2_{_{NP}})-\gamma_{_E},\;$ $\frac{1}{\varepsilon}$ is an infinite term, the mass scale $\kappa$ is introduced in the dimensional regularization, $\Lambda_{NP}$ is the NP scale, and $ \gamma_{E} $ is the Euler-Mascheroni constant. $x_i$ represents $\frac{m_i^2}{\Lambda_{NP}^2}$, and the concrete form of formula $F_{21}(x_{1},x_{2},x_{3})$ is given in the Appendix.

We can see that the infinite terms of the WCs of FIG.~\ref{fig1} (a) are $\mathcal{C}_{VL(a)}^{\ell (IF)}=\frac{g_L^{2}}{16 \pi^{2}}\Delta_{UV}$ and $\mathcal{C}_{AL(a)}^{\ell (IF)}=-\mathcal{C}_{VL(a)}^{\ell (IF)}$. Similarly, the infinite terms of the WCs of the following three diagrams (FIG.~\ref{fig1} (b),(c) and (d)) are given as follows:
\begin{eqnarray}
\mathcal{C}_{VL(b)}^{\ell (IF)}&=&\frac{e^2}{64\pi^{2}s_{W}^{2}c_{W}^{2}}(1-2c_{W}^{2})\Delta_{UV},~~~
\mathcal{C}_{AL(b)}^{\ell (IF)}~=-\mathcal{C}_{VL(b)}^{\ell},\nonumber\\&&\hspace{-1.8cm}
\mathcal{C}_{VL(c)}^{\ell (IF)}~=\frac{e^{2}}{32\pi^{2}s_{W}^{2}}\Delta_{UV},~~~~~~~~~~~~~~~~~~~~
\mathcal{C}_{AL(c)}^{\ell (IF)}~=-\mathcal{C}_{VL(c)}^{\ell},\nonumber\\&&\hspace{-1.8cm}
\mathcal{C}_{VL(d)}^{\ell (IF)}~=(\frac{e^{2}}{32\pi^{2}s_{W}^{2}}+\frac{(Y_{l}^{\ell})^{2}}{32\pi^{2}})\Delta_{UV},~~~~~~
\mathcal{C}_{AL(d)}^{\ell (IF)}~=-\mathcal{C}_{VL(d)}^{\ell}.
\end{eqnarray}
Now we should deal with the UV divergences by renormalization procedures.
\subsubsection{The counter term in the on-shell scheme}

Considering the final state lepton and neutrino are both real particles, we use the on-shell scheme to eliminate the infinite terms.
To obtain finite results, the contributions from the counter terms for the vertex $\overline{l^I}\nu^IW^-$ are necessary. The counter term formula for the vertex $\overline{l^I}\nu^IW^-$ is:
\begin{eqnarray}
&&\delta V_{\overline{l^I}\nu^IW^-}^{\mu~(OS)}=
\frac{-i e}{2\sqrt{2}s_{W}}\Big(\frac{\delta
m_{Z}^2}{m_{Z}^2}-\frac{\delta m_{Z}^2 -\delta m_{W}^2}{m_{Z}^2-m_{W}^2}+2\delta
e+{\delta Z_{L}^{l}}^{I}+{\delta Z_{L}^{\nu}}^{I}+\delta
Z_{WW}\Big)\gamma^\mu P_{L},
\end{eqnarray}
Following the method in Refs. \cite{onshell1,onshell2,onshell3}, we obtain the needed
renormalization constants in the BLMSSM.

We calculate the Z boson self-energy diagram (loop particles are sneutrinos or sleptons) and get the renormalization constant $\delta
m_{Z}^2$:
\begin{eqnarray}
&&\frac{\delta
m_{Z}^2}{m_{Z}^2}=[\frac{e^2}{32\pi^2s_{W}^2c_{W}^2}(1-2s_{W}^2)^2+\frac{e^2}{16 \pi^2 c_{W}^2}]\Delta_{UV}
-\frac{e^2}{s_{W}^2c_{W}^2}\Big\{\frac{1}{4}\sum_{j=1}^6F_1(x_{\tilde{\nu}_{j}},x_{\tilde{\nu}_{j}})
\nonumber\\&&\hspace{1.5cm}+\sum_{\alpha,\beta=1}^6|(\mathcal{G})_{\alpha\beta}|^2
F_1(x_{\tilde{L}_{\alpha}},x_{\tilde{L}_{\beta}})\Big\}.
\end{eqnarray}
Through calculating the W boson self-energy diagram (with sneutrinos and sleptons in the loop), we can obtain:
\begin{eqnarray}
&& \frac{\delta
m_{W}^2}{m_{Z}^2}=\frac{e^2c_{W}^2}{32\pi^2s_{W}^2}\Delta_{UV}
-\frac{e^2c_{W}^2}{2s_{W}^2}\sum_{i=1}^6\sum_{\alpha=1}^6
|(\mathcal{\eta})_{i\alpha}|^2F_1(x_{\tilde{\nu}_{\alpha}},x_{\tilde{L}_{i}}),\nonumber\\
&&\delta
Z_{WW}=-\frac{e^2}{32\pi^2s_{W}^2}\Delta_{UV}+\frac{e^2}{2s_{_W}^2}\sum_{i=1}^6\sum_{\alpha=1}^6
|(\mathcal{\eta})_{i\alpha}|^2F_1(x_{\tilde{\nu}_{\alpha}},x_{\tilde{L}_{i}}).
\end{eqnarray}
The renormalization constant of charge is obtained from virtual sleptons:
\begin{eqnarray}
&&\delta
e=\frac{e^2}{16\pi^2}\Delta_{UV}-\frac{1}{2} e^2\sum_{i=1}^6F_1(x_{\tilde{L}_{i}},x_{\tilde{L}_{i}}).
\end{eqnarray}
In the same way, we give the renormalization constants ${\delta Z_L^{\nu}}^{I}$ and ${\delta
Z_L^l}^{I}$ for neutrinos and leptons respectively:
\begin{eqnarray}
&&{\delta Z_L^{\nu}}^{I}=\frac{-e^2}{32\pi^2s_{W}^2}\Big(\frac{1}{2c_{W}^2}+1
+(\frac{s_{W}Y_{l}^{I}}{e})^2+(\frac{\sqrt{2}s_{W}g_{L}}{e})^2\Big)\Delta_{UV}\nonumber\\&&\hspace{1.5cm}
-\frac{e^2}{2s^2_{W}c^2_{W}}\sum_{i=1}^4\sum_{\alpha=1}^6|(\zeta^I)_{\alpha i}|^2
F_2(x_{\tilde{\nu}_{\alpha}},x_{\chi_{i}^0})-\frac{e^2}{s^2_{W}}\sum_{i=1}^2\sum_{\alpha=1}^6|(\mathcal{P}^I)_{\alpha i}|^2
F_2(x_{\tilde{L}_{\alpha}},x_{\chi_{i}^-})\nonumber\\&&\hspace{1.5cm}
-\frac{e^2}{2s^2_{W}c^2_{_W}}\sum_{i=1}^3\sum_{\alpha=1}^6|(\zeta^{\prime I})_{\alpha i}|^2
F_2(x_{\tilde{\nu}_{\alpha}},x_{\chi_{Li}^0}),\nonumber\\
&&{\delta Z_L^l}^{I}=\frac{-e^2}{32\pi^2s_{W}^2}\Big(\frac{1}{2c_{W}^2}+1
+(\frac{s_{W}Y_{l}^{I}}{e})^2+(\frac{\sqrt{2}s_{W}g_{L}}{e})^2\Big)\Delta_{UV}\nonumber\\&&\hspace{1.5cm}
-\frac{e^2}{s_{W}^2}\sum_{\alpha=1}^6\sum_{i=1}^2\Big\{|(\mathcal{B}_i)^{I\alpha}|^2
F_2+x_{e^I}\Big[|(\mathcal{B}_i)^{I\alpha}|^2\nonumber\\&&\hspace{1.5cm}+|(\mathcal{A}_i)^{I\alpha}|^2+
2\mathbf{Re}[(\mathcal{A}_i^{\dag})^{I\alpha}(\mathcal{B}_i)^{I\alpha}]\Big]
F_3\Big\}(x_{\tilde{\nu}_{\alpha}},x_{\chi_i^-})\nonumber\\&&\hspace{1.5cm}
-e^2\sum_{j=1}^4\sum_{i=1}^6\Big\{x_{e^I}\!\Big[\frac{|(\mathcal{D}^I)_{ij}|^2}{2s^2_{W}}\!
+\!\frac{\sqrt{2}}{s_{W}}\mathbf{Re}[(\mathcal{C}^I)^{\dag}_{ij}
(\mathcal{D}^I)_{ij}]\nonumber\\&&\hspace{1.5cm}+|(\mathcal{C}^I)_{ij}|^2\Big]F_3
+\frac{1}{2s^2_{W}}|(\mathcal{D}^I)_{ij}|^2F_2\Big\}(x_{\tilde{L}_{i}},x_{\chi_j^0})\nonumber\\&&\hspace{1.5cm}
-e^2\sum_{j=1}^3\sum_{i=1}^6\Big\{x_{e^I}\!\Big[\frac{|(\mathcal{D}^{\prime I})_{ij}|^2}{2s^2_{W}}\!
+\!\frac{\sqrt{2}}{s_{W}}\mathbf{Re}[(\mathcal{C}^{\prime I})^{\dag}_{ij}
(\mathcal{D}^{\prime I})_{ij}]\nonumber\\&&\hspace{1.5cm}+|(\mathcal{C}^{\prime I})_{ij}|^2\Big]F_3
+\frac{1}{2s^2_{W}}|(\mathcal{D}^{\prime I})_{ij}|^2F_2\Big\}(x_{\tilde{L}_{i}},x_{\chi_{Lj}^0}),
\end{eqnarray}
where the vertex couplings are given by
\begin{eqnarray}
&&(\mathcal{A}_i)^{I\alpha}={Y_{l}^{I}s_{W}\over e}Z_{-}^{2i*}
Z^{I\alpha*}_{\tilde{\nu}},~~~
(\mathcal{B}_i)^{I\alpha}=Z_{+}^{1i}Z^{I\alpha*}_{\tilde{\nu}},~~~(\mathcal{\eta})_{i\alpha}=Z_{\tilde{\nu}}^{J\alpha}Z_{\tilde{_L}}^{Ji},~~
\nonumber\\&&
(\mathcal{P}^I)_{\alpha i}=\frac{-Y_{l}^{J}s_{W}}{e}
Z_{\nu}^{JI}Z_{\tilde{L}}^{(J+3)\alpha*}Z^{2i*}_{-}-Z_{\nu}^{JI}Z_{\tilde{L}}^{J\alpha*}Z^{1i*}_{-},
\nonumber\\&&(\mathcal{\zeta}^{I})_{\alpha i}=Z_{\tilde{\nu}}^{J\alpha*}Z_{\nu}^{JI}
(Z_{N}^{1i}s_{W}-Z_{N}^{2i}c_{W}),
~~~(\mathcal{C}^I)_{ij}=\frac{-\sqrt{2}}{c_{W}}Z_{\tilde{L}}^{(I+3)i}Z_{N}^{1j*}
+\frac{Y_{l}^{I}}{e}Z_{\tilde{L}}^{Ii}Z_{N}^{3j*},\nonumber\\
&&(\mathcal{D}^I)_{ij}=\frac{Z_{\tilde{L}}^{Ii}}{c_{W}}
(\!Z_{N}^{1j}s_{W}\!+\!Z_{N}^{2j}c_{W}\!)
\!+\!\frac{\sqrt{2}s_{W}Y_{l}^{I}}{e}Z_{\tilde{L}}^{(I+3)i}
Z_{N}^{3j},\nonumber\\
&&(\mathcal{G})_{\alpha\beta}=\frac{1}{2}Z_{\tilde{L}}^{J\alpha}
Z_{\tilde{L}}^{J\beta*}-s^2_{W}\delta^{\alpha\beta},
~~~(\mathcal{C}^{\prime I})_{ij}=\frac{-\sqrt{2}g_{L}}{e}Z_{NL}^{1j*}Z_{L}^{(I+3)i},~~~(\mathcal{D}^{\prime I})_{ij}=\frac{2g_{L}s_{W}}{e}
Z_{NL}^{1j}Z_{L}^{Ii},
\nonumber\\&&(\mathcal{\zeta}^{\prime I})_{\alpha i}=\frac{\sqrt{2} s_{W} c_{W}}{e} (\sqrt{2}g_{L} Z_{NL}^{1i} Z_{\nu}^{JI} Z_{\tilde{\nu}}^{J\alpha*}-[2\lambda_{N^{c}}^{JJ} Z_{NL}^{3i}+\sqrt{2}g_{L} Z_{NL}^{1i}] Z_{\nu}^{(J+3)I} Z_{\tilde{\nu}}^{(J+3) \alpha *} ).
\end{eqnarray}
The functions $F_1, F_2$ and $F_3$ are as follows:
\begin{eqnarray}
&&\hspace{-1.6cm}F_1(x_{1},x_{2})=\frac{1}{288\pi^2(x_{1}-x_{2})^3}[6(x_{1}-3x_{2})x_{1}^2\ln
x_{1}\!+\!6(3x_{1}-x_{2})x_{2}^2\ln x_{2} \nonumber\\&&\hspace{0.8cm}\!\! -(x_{1}-x_{2})(5x_{1}^2-22x_{1}x_{2}+5x_{2}^2)],\nonumber\\&&\hspace{-1.6cm}
F_2(x_{1},x_{2})=\frac{(2x_{2}-x_{1})(x_{2}-x_{1}+x_{1}\ln x_{1})-x_{2}^2\ln x_{2} }{32\pi^2(x_{1}-x_{2})^2},\nonumber\\&&\hspace{-1.6cm}
F_3(x_{1},x_{2})=\frac{x_{1}^2+2x_{1}x_{2}(\ln
x_{2}-\ln x_{1})-x_{2}^2}{32\pi^2(x_{1}-x_{2})^3}.
\end{eqnarray}
If $x_{1}=x_{2}$, they simplify to
\begin{eqnarray}
&&\hspace{-1.6cm}F_1(x_{1},x_{2})=\frac{\ln x_{1}}{48\pi^2},~~~
F_2(x_{1},x_{2})=-\frac{\ln x_{1}}{32\pi^2}+\frac{1}{64\pi^2},~~~
F_3(x_{1},x_{2})=\frac{1}{96\pi^2x_{1}}.
\end{eqnarray}

Now, the WCs of FIG.~\ref{fig1} (counter) read:
\begin{eqnarray}
{\mathcal{C}_{VL}^{\ell}}_{(counter)}&=&\frac{1}{2}\Big(\frac{\delta
m_{Z}^2}{m_{Z}^2}-\frac{\delta m_{Z}^2 -\delta m_{W}^2}{m_{Z}^2-m_{W}^2}+2\delta
e+{\delta Z_{L}^{l}}^{\ell}+{\delta Z_{L}^{\nu}}^{\ell}+\delta
Z_{WW}\Big),\nonumber\\&&\hspace{-2.68cm}
{\mathcal{C}_{AL}^{\ell}}_{(counter)}~=-{\mathcal{C}_{VL}^{\ell}}_{(counter)}.
\end{eqnarray}
The corresponding $\mathcal{C}_{VL(counter)}^{\ell (IF)}$ and $\mathcal{C}_{AL(counter)}^{\ell (IF)}$ are:
\begin{eqnarray}
\mathcal{C}_{VL(counter)}^{\ell (IF)}&=&\frac{1}{2}\Big\{\Big[\frac{e^2}{32\pi^2s_{W}^2c_{W}^2}(1-2s_{W}^2)^2+\frac{e^2}{16 \pi^2 c_{W}^2}\Big]\Delta_{UV}\nonumber\\&&\hspace{-0.2cm}-\Big[\Big(\frac{e^2}{32\pi^2s_{W}^2c_{W}^2}(1-2s_{W}^2)^2+\frac{e^2}{16 \pi^2 c_{W}^2}\Big)\Delta_{UV}-\frac{e^2c_{W}^2}{32\pi^2s_{W}^2}\Delta_{UV}\Big]\frac{m_{Z}^{2}}{m_{Z}^{2}-m_{W}^{2}}\nonumber\\&&\hspace{-0.2cm}+
\frac{e^2}{8\pi^2}\Delta_{UV}-\frac{e^2}{32\pi^2s_{W}^2}\Delta_{UV}+\frac{-e^2}{16\pi^2s_{W}^2}\Big[\frac{1}{2c_{W}^2}+1 \nonumber\\&&\hspace{-0.2cm}
+(\frac{s_{W}Y_{l}^{\ell}}{e})^2+(\frac{\sqrt{2}s_{W}g_{L}}{e})^2\Big]\Delta_{UV}\Big\},\nonumber\\&&\hspace{-2.58cm}
\mathcal{C}_{AL(counter)}^{\ell (IF)}~=-\mathcal{C}_{VL(counter)}^{\ell (IF)}.
\end{eqnarray}

It is easy to test that the infinite terms in the sum of FIG.~\ref{fig1} (a),(b),(c),(d) and (counter) vanish: $\mathcal{C}_{VL}^{\ell (IF)}=\mathcal{C}_{VL(a)}^{\ell (IF)}+\mathcal{C}_{VL(b)}^{\ell (IF)}+\mathcal{C}_{VL(c)}^{\ell (IF)}+\mathcal{C}_{VL(d)}^{\ell (IF)}+\mathcal{C}_{VL(counter)}^{\ell (IF)}=0$, similarly, $\mathcal{C}_{AL}^{\ell (IF)}=0$. Therefore, the divergences are completely eliminated.
Note that the infinite terms in the sum of FIG.~\ref{fig1} (a),(b),(c) and (d) can be eliminated by the counter terms.
However, a single diagram in FIG.~\ref{fig1} (b),(c),(d), such as FIG.~\ref{fig1} (b), cannot be counteracted individually in the on-shell scheme.
\subsection{Box-type Feynman diagrams}
\begin{figure}[!htbp]
\centering
\includegraphics[width=12cm]{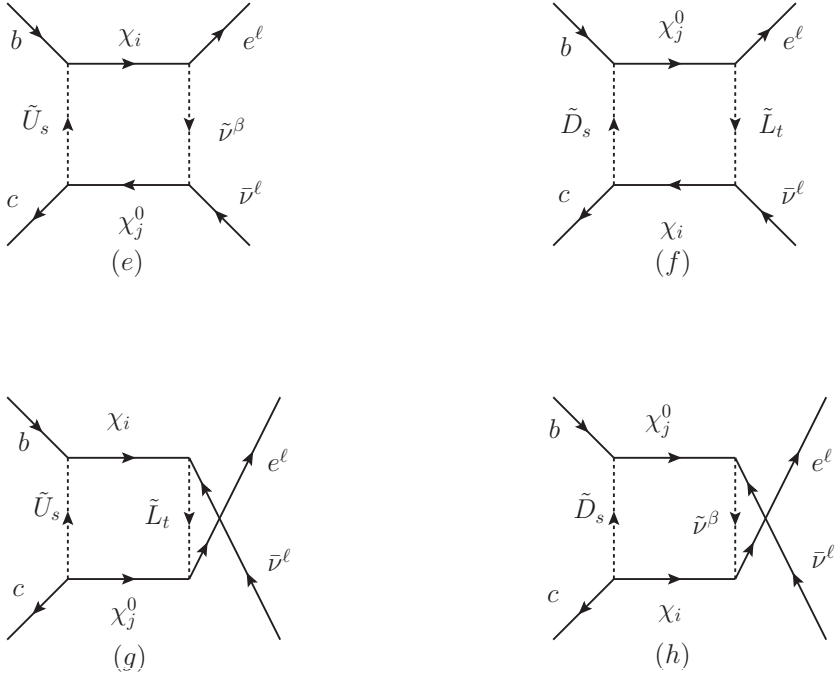}\\
\caption{The box-type Feynman diagrams that can correct $R_{D^{(*)}}$ in BLMSSM} \label{fig2}
\end{figure}
Taking FIG.~\ref{fig2} (e) as an example, the corresponding WCs are given as follows:
\begin{eqnarray}
\mathcal{C}_{VL(e)}^{\ell}&=&\sum_{\beta,s=1}^{6}\sum_{i=1}^{2}\sum_{j=1}^{4}
[\frac{\mathcal{B}_{1}^{\beta\ell i}\mathcal{B}_{2}^{i s}\mathcal{A}_{3}^{s j}\mathcal{A}_{4}^{\beta\ell j}m_{\chi^0_{j}}m_{\chi_{i}}}{\Lambda_{NP}^{4}}\frac{1}{64\pi^{2}}
F_{11}(x_{\tilde{\nu}^{\beta}},
x_{\chi^0_{j}},x_{\tilde{U}_{s}},x_{\chi_{i}})\nonumber\\&&
-\frac{\mathcal{B}_{1}^{\beta\ell i}\mathcal{A}_{2}^{i s}\mathcal{B}_{3}^{s j}\mathcal{A}_{4}^{\beta\ell j}}{\Lambda_{NP}^{2}}\frac{1}{128\pi^{2}}F_{21}(x_{\tilde{\nu}^{\beta}},
x_{\chi^0_{j}},x_{\tilde{U}_{s}},x_{\chi_{i}})]/(\sqrt{2}G_{F}V_{cb}),
\label{CVLe}
\end{eqnarray}
\begin{eqnarray}
\mathcal{C}_{AL(e)}^{\ell}&=&\sum_{\beta,s=1}^{6}\sum_{i=1}^{2}\sum_{j=1}^{4}[\frac{\mathcal{B}_{1}^{\beta\ell i}\mathcal{B}_{2}^{i s}\mathcal{A}_{3}^{s j}\mathcal{A}_{4}^{\beta\ell j}m_{\chi^0_{j}}m_{\chi_{i}}}{\Lambda_{NP}^{4}}\frac{1}{64\pi^{2}}F_{11}(x_{\tilde{\nu}^{\beta}},
x_{\chi^0_{j}},x_{\tilde{U}_{s}},x_{\chi_{i}})\nonumber\\&&\hspace{0cm}
+\frac{\mathcal{B}_{1}^{\beta\ell i}\mathcal{A}_{2}^{i s}\mathcal{B}_{3}^{s j}\mathcal{A}_{4}^{\beta\ell j}}{\Lambda_{NP}^{2}}\frac{1}{128\pi^{2}}F_{21}
(x_{\tilde{\nu}^{\beta}},
x_{\chi^0_{j}},x_{\tilde{U}_{s}},x_{\chi_{i}})]/(\sqrt{2}G_{F}V_{cb}),
\label{CALe}
\end{eqnarray}
\begin{eqnarray}
\mathcal{C}_{SL(e)}^{\ell}&=&\sum_{\beta,s=1}^{6}\sum_{i=1}^{2}\sum_{j=1}^{4}[\frac{\mathcal{A}_{1}^{\beta\ell i}\mathcal{A}_{2}^{i s}\mathcal{A}_{3}^{s j}\mathcal{A}_{4}^{\beta\ell j}m_{\chi^0_{j}}m_{\chi_{i}}}{\Lambda_{NP}^{4}}\frac{1}{64\pi^{2}}F_{11}(x_{\tilde{\nu}^{\beta}},
x_{\chi^0_{j}},x_{\tilde{U}_{s}},x_{\chi_{i}})\nonumber\\&&\hspace{0cm}
+\frac{\mathcal{A}_{1}^{\beta\ell i}\mathcal{B}_{2}^{i s}\mathcal{B}_{3}^{s j}\mathcal{A}_{4}^{\beta\ell j}}{\Lambda_{NP}^{2}}
\frac{1}{64\pi^{2}}F_{21}(x_{\tilde{\nu}^{\beta}},
x_{\chi^0_{j}},x_{\tilde{U}_{s}},x_{\chi_{i}})]/(\sqrt{2}G_{F}V_{cb}),
\label{CSLe}
\end{eqnarray}
\begin{eqnarray}
\mathcal{C}_{TL(e)}^{\ell}&=&[\sum_{\beta,s=1}^{6}\sum_{i=1}^{2}\sum_{j=1}^{4}
\frac{\mathcal{A}_{1}^{\beta\ell i}\mathcal{A}_{2}^{i s}\mathcal{A}_{3}^{s j}\mathcal{A}_{4}^{\beta\ell j}m_{\chi^0_{j}}m_{\chi_{i}}}{\Lambda_{NP}^{4}}
\frac{1}{128\pi^{2}}F_{11}(x_{\tilde{\nu}^{\beta}},
x_{\chi^0_{j}},x_{\tilde{U}_{s}},x_{\chi_{i}})]\nonumber\\&&\hspace{0cm}/(\sqrt{2}G_{F}V_{cb}),
\label{CTLe}
\end{eqnarray}
\begin{eqnarray}
\mathcal{C}_{PL(e)}^{\ell}&=&\sum_{\beta,s=1}^{6}\sum_{i=1}^{2}\sum_{j=1}^{4}[-\frac{(\mathcal{A}_{1}^{\beta\ell i}\mathcal{A}_{2}^{i s}-\mathcal{B}_{1}^{\beta\ell i}\mathcal{B}_{2}^{i s})\mathcal{A}_{3}^{s j}\mathcal{A}_{4}^{\beta\ell j}m_{\chi^0_{j}}m_{\chi_{i}}}{\Lambda_{NP}^{4}}\frac{1}{128\pi^{2}}F_{11}(x_{\tilde{\nu}^{\beta}},
x_{\chi^0_{j}},x_{\tilde{U}_{s}},x_{\chi_{i}})\nonumber\\&&\hspace{0cm}
+\frac{\mathcal{A}_{1}^{\beta\ell i}\mathcal{B}_{2}^{i s}\mathcal{B}_{3}^{s j}\mathcal{A}_{4}^{\beta\ell j}}{\Lambda_{NP}^{2}}\frac{1}{64\pi^{2}}
F_{21}(x_{\tilde{\nu}^{\beta}},
x_{\chi^0_{j}},x_{\tilde{U}_{s}},x_{\chi_{i}})]/(\sqrt{2}G_{F}V_{cb}),
\label{CPLe}
\end{eqnarray}
\begin{eqnarray}
\mathcal{C}_{PR(e)}^{\ell}&=&[\sum_{\beta,s=1}^{6}\sum_{i=1}^{2}\sum_{j=1}^{4}-\frac{(\mathcal{A}_{1}^{\beta\ell i}\mathcal{A}_{2}^{i s}+\mathcal{B}_{1}^{\beta\ell i}\mathcal{B}_{2}^{i s})\mathcal{A}_{3}^{s j}\mathcal{A}_{4}^{\beta\ell j}m_{\chi^0_{j}}m_{\chi_{i}}}{\Lambda_{NP}^{4}}\frac{1}{128\pi^{2}}F_{11}(x_{\tilde{\nu}^{\beta}},
x_{\chi^0_{j}},x_{\tilde{U}_{s}},x_{\chi_{i}})]\nonumber\\&&\hspace{0cm}
/(\sqrt{2}G_{F}V_{cb}),
\label{CPRe}
\end{eqnarray}
and all the other WCs in Eq. (\ref{EL}) vanish. In Eqs. (\ref{CVLe}-\ref{CPRe}),
\begin{eqnarray}
&&\mathcal{A}_{1}^{\beta\ell i}=-Y_{l}^{\ell}Z_{-}^{2i}Z_{\tilde{\nu}}^{\ell\beta},~~~
\mathcal{A}_{2}^{i s}=\sum_{I=1}^{3}(\frac{-e}{s_{W}}Z_{\tilde{U}}^{Is*}Z_{+}^{1i}+Y_{u}^{I}Z_{\tilde{U}}^{(I+3)s*}Z_{+}^{2i})V_{I3},\nonumber\\&&\hspace{0cm}
\mathcal{A}_{3}^{s j}=\frac{2\sqrt{2}e}{3c_{W}}Z_{\tilde{U}}^{5s}Z_{N}^{1j}-Y_{u}^{2}Z_{\tilde{U}}^{2s}Z_{N}^{4j},\nonumber\\&&\hspace{0cm}
\mathcal{A}_{4}^{\beta\ell j}=\sum_{I=1}^{3}\{Z_{\tilde{\nu}}^{I\beta*}Z_{\nu}^{I\ell}\frac{e}{\sqrt{2}s_{W}c_{W}}(Z_{N}^{1j}s_{W}-Z_{N}^{2j}c_{W})+\sum_{J=1}^{3}(\frac{Y_{\nu}^{IJ}}{\sqrt{2}}Z_{N}^{4j}(Z_{\nu}^{I\ell}Z_{\tilde{\nu}}^{(J+3)\beta*}+Z_{\nu}^{(I+3)\ell}Z_{\tilde{\nu}}^{J\beta*}))\},\nonumber\\&&\hspace{0cm}
\mathcal{B}_{1}^{\beta\ell i}=-\frac{e}{s_{W}}Z_{+}^{1i*}Z_{\tilde{\nu}}^{\ell\beta}-\sum_{I=1}^{3}(Y_{\nu}^{\ell I}Z_{+}^{2i*}Z_{\tilde{\nu}}^{(I+3)\beta}),~~~
\mathcal{B}_{2}^{i s}=-\sum_{I=1}^{3}(Y_{d}^{3}Z_{\tilde{U}}^{Is*}Z_{-}^{2i*})V_{I3},\nonumber\\&&\hspace{0cm}
\mathcal{B}_{3}^{s j}=-\frac{e}{\sqrt{2}s_{W}c_{W}}Z_{\tilde{U}}^{2s}(\frac{1}{3}Z_{N}^{1j*}s_{W}+Z_{N}^{2j*}c_{W})-Y_{u}^{2}Z_{\tilde{U}}^{5s}Z_{N}^{4j*}.
\end{eqnarray}
The formulae $F_{11}(x_{1},x_{2},x_{3},x_{4})$ and $F_{21}(x_{1},x_{2},x_{3},x_{4})$ are given in the Appendix.
\section{Numerical results}
For the numerical discussion, the parameters used are:\\
$m_{1}=400 \rm{GeV}$,~~~ $M_{L}=2000 \rm{GeV}$,~~~ $\mu_{L}=1600 \rm{GeV}$,\\
$\rm{tan}\beta_{L}=0.1$,~~~ $v_{L}=1260 \rm{GeV}$,~~~ $\lambda_{N^{c}}=1$,~~~ $\Lambda_{NP}=1000 \rm{GeV}$,\\
$(m_{\tilde{Q}}^{2})_{ii}=(m_{\tilde{U}}^{2})_{ii}=(m_{\tilde{D}}^{2})_{ii}=(m_{\tilde{N}^{c}}^{2})_{ii}=3\times10^{6} \rm{GeV^{2}}$,\\
$(A_{l})_{ii}=(A_{l}^{'})_{ii}=300 \rm{GeV}$,~~~ $(A_{N})_{ii}=(A_{N^{c}})_{ii}=500 \rm{GeV}$,\\
and $(A_{u})_{ii}=(A_{d})_{ii}=(A_{u}^{'})_{ii}=(A_{d}^{'})_{ii}=500 \rm{GeV}$,\\
where $i=1\dots 3$. If not otherwise noted, the non-diagonal elements of the parameters used should be zero. The Yukawa couplings of neutrinos $Y_{\nu}^{I J}$ are of the order of $10^{-8}\sim10^{-6}$; their effects are tiny and can be ignored.

At present, all supersymmetric mass bounds are model-dependent. Based on the PDG\cite{PDG} data, we consider the limitations on masses of the charginos and neutralinos (the strongest limitations are $345\rm{GeV}$). In our work, the masses of charginos $m_{\chi^{\pm}}\simeq(1000\sim2000)\rm{GeV}$ and the masses of neutralinos $m_{\chi^{0}}\simeq(400\sim2000)\rm{GeV}$, all of which can satisfy the mass bounds. The limits for the sleptons are around $290\rm{GeV}\sim 450\rm{GeV}$ \cite{LHCslepton}, which can be satisfied easily. The masses of squarks in this paper are larger than $1000\rm{GeV}$, so the limits for squarks are also satisfied. In other words, the parameters given above and the parameter space to be discussed below can all satisfy the mass bounds.
\subsection{Effects of parameters $m_{\tilde{L}}^{2}$ (or $m_{\tilde{R}}^{2}$) on $R_{D^{(*)}}$}
\begin{figure}[!htbp]
\begin{center}
\begin{minipage}[c]{0.48\textwidth}
\includegraphics[width=3in]{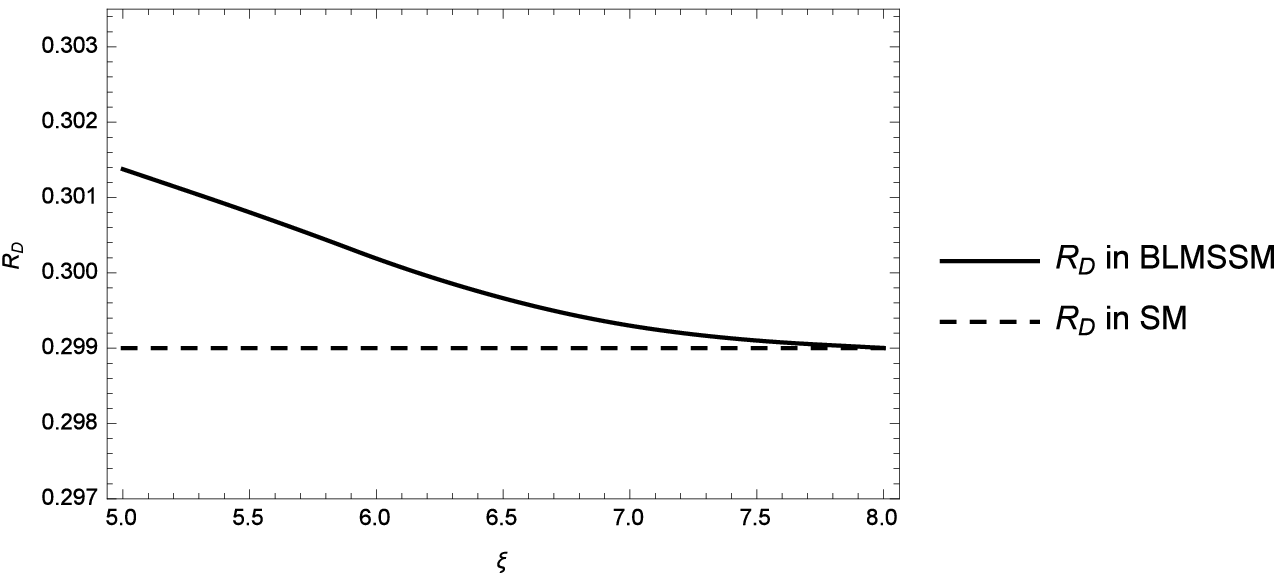}
\end{minipage}%
\begin{minipage}[c]{0.48\textwidth}
\includegraphics[width=3in]{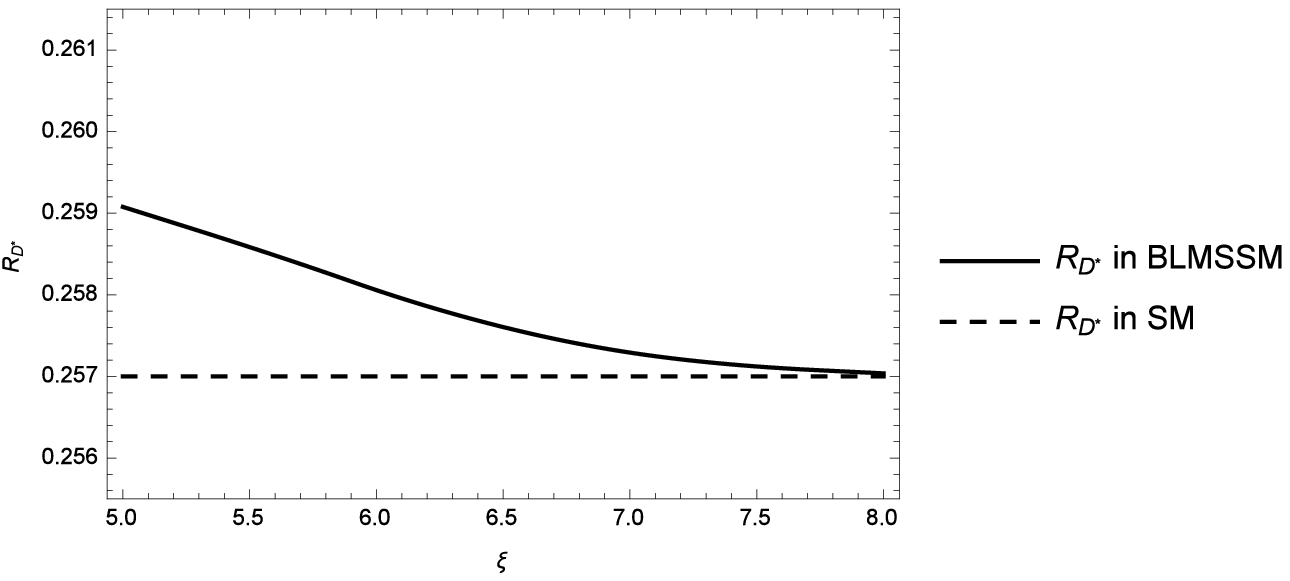}
\end{minipage}
\caption[]{\label{fig3} With $(m_{\tilde{L}}^{2})_{11}=(m_{\tilde{R}}^{2})_{11}=(m_{\tilde{L}}^{2})_{22}=(m_{\tilde{R}}^{2})_{22}=3\times 10^{\xi} ~\rm{GeV^{2}}$, and $(m_{\tilde{L}}^{2})_{33}=(m_{\tilde{R}}^{2})_{33}=3\times10^{8} \rm{GeV^{2}}$, the results versus $\xi$ are plotted.}
\end{center}
\end{figure}

We now focus on the effects of parameters $m_{\tilde{L}}^{2}$ (or $m_{\tilde{R}}^{2}$) on $R_{D^{(*)}}$. First, we set the parameters as follows:\\
$\rm{tan}\beta=10$,~~~ $m_{2}=\mu=1200 \rm{GeV}$,~~~ $g_{L}=0.1$,~~~and $(m_{\tilde{L}}^{2})_{33}=(m_{\tilde{R}}^{2})_{33}=3\times10^{8} \rm{GeV^{2}}$.\\
To study the impacts of these parameters on $R_{D^{(*)}}$, we used the parameters $(m_{\tilde{L}}^{2})_{11}=(m_{\tilde{R}}^{2})_{11}=(m_{\tilde{L}}^{2})_{22}=(m_{\tilde{R}}^{2})_{22}=3\times 10^{\xi} ~\rm{GeV^{2}}$, where ${\xi}$ is a variable. After calculation we obtain FIG.~\ref{fig3}. Here, we use the central value of the SM prediction in our calculation. The left-hand diagram shows $R_{D}$ and the right-hand diagram shows $R_{D^{*}}$.

We know the measurement of $R_{D^{(*)}e}$ (which implies $l=e$ in Eq. (\ref{DFRDX})) is approximately equal to that of $R_{D^{(*)}\mu}$ ($l=\mu$ in Eq. (\ref{DFRDX})).
This is the reason why we set $(m_{\tilde{L}}^{2})_{11}=(m_{\tilde{R}}^{2})_{11}=(m_{\tilde{L}}^{2})_{22}=(m_{\tilde{R}}^{2})_{22}$.
To solve the problem of $R_{D^{(*)}}$, we should violate lepton flavour symmetry for generations 1(2) and generation 3. Therefore, we suppose $(m_{\tilde{L}}^{2})_{33}=(m_{\tilde{R}}^{2})_{33}\neq(m_{\tilde{L}}^{2})_{11}$.

It is easy to see from FIG.~\ref{fig3} that $R_{D^{(*)}}$ decreases as ${\xi}$ increases. Obviously, our results satisfy the decoupling rule. When the
sleptons are very heavy, the BLMSSM results are very near the SM predictions. In fact, the SM predictions of $R_{D^{(*)}}$ cannot explain experimental values well,
and our goal is to increase the theoretical values. From the numerical analysis, the following relational expression should be set up: $(m_{\tilde{L}}^{2})_{11}=(m_{\tilde{R}}^{2})_{11}=(m_{\tilde{L}}^{2})_{22}=(m_{\tilde{R}}^{2})_{22}<(m_{\tilde{L}}^{2})_{33}=(m_{\tilde{R}}^{2})_{33}$. We need to select a set of reasonable parameters, and finally choose: $(m_{\tilde{L}}^{2})_{11}=(m_{\tilde{R}}^{2})_{11}=(m_{\tilde{L}}^{2})_{22}=(m_{\tilde{R}}^{2})_{22}=5.5\times10^{5} \rm{GeV^{2}}$, and $(m_{\tilde{L}}^{2})_{33}=(m_{\tilde{R}}^{2})_{33}=3\times10^{8} \rm{GeV^{2}}$.
Up to now, our theoretical values of $R_{D^{(*)}}$ are only a little bigger than those of the SM,
so we also need to study the effects of other parameters on $R_{D^{(*)}}$.
\subsection{Effect of parameter $g_{L}$ on $R_{D^{(*)}}$}
Based on the above analysis, we use the following parameters:\\
$\rm{tan}\beta=10$,~~~ $m_{2}=\mu=1200 \rm{GeV}$,\\$(m_{\tilde{L}}^{2})_{11}=(m_{\tilde{R}}^{2})_{11}=(m_{\tilde{L}}^{2})_{22}=(m_{\tilde{R}}^{2})_{22}=5.5\times10^{5} \rm{GeV^{2}}$, and $(m_{\tilde{L}}^{2})_{33}=(m_{\tilde{R}}^{2})_{33}=3\times10^{8} \rm{GeV^{2}}$.\\

\begin{figure}[!htbp]
\begin{center}
\begin{minipage}[c]{0.48\textwidth}
\includegraphics[width=3in]{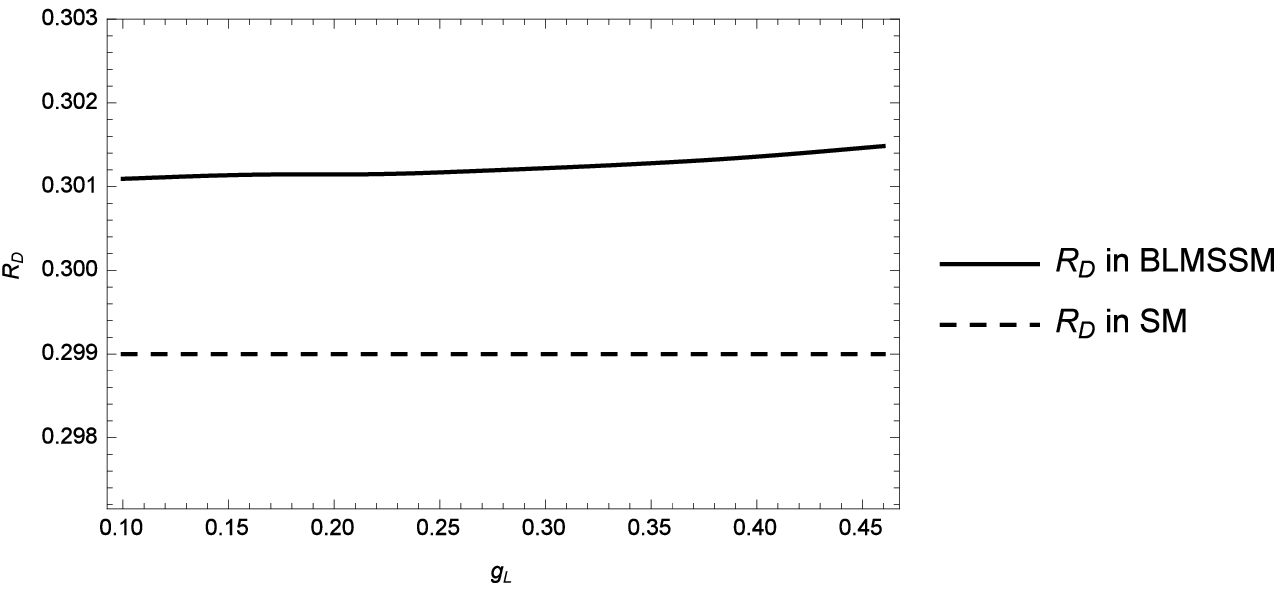}
\end{minipage}%
\begin{minipage}[c]{0.48\textwidth}
\includegraphics[width=3in]{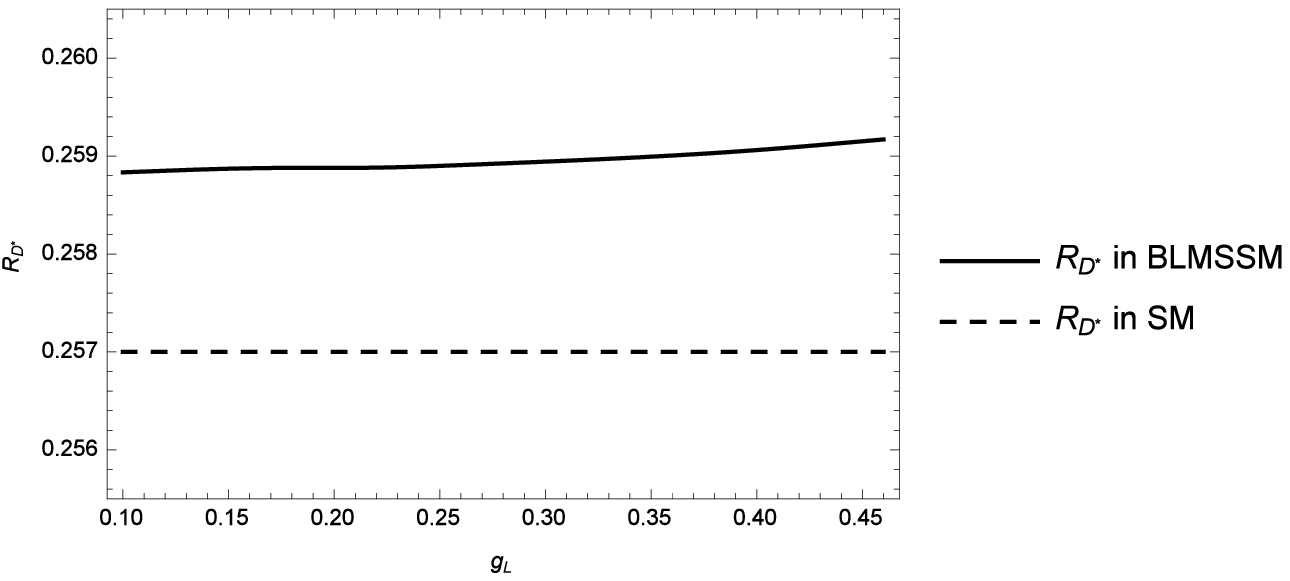}
\end{minipage}
\caption[]{\label{fig4} The diagrams of $R_{D}$ (left) and $R_{D^{*}}$ (right) versus $g_L$.}
\end{center}
\end{figure}

$g_{L}$ is the coupling constant of the vertexes $l \chi_L^0\tilde{L}$ and $\nu \chi_L^0\tilde{\nu}$.
 As a new parameter in the BLMSSM, $g_{L}$ should affect $R_{D^{(*)}}$, which is of interest.
The obtained numerical results are plotted in FIG.~\ref{fig4}. The left-hand diagram shows $R_{D}$ and the right-hand diagram shows $R_{D^{*}}$.

From FIG.~\ref{fig4}, we can see that $R_{D}$ and $R_{D^{*}}$ both increase gently
 with increasing $g_L$. This is easy to understand: larger $g_L$ improves the effects from NP. In order to get larger theoretical values of $R_{D^{(*)}}$,
  we need to choose a larger $g_L$. After considering the reasonableness of the range of parameter $g_L$,
    we use $g_L=0.45$. In this case, our numerical results are further improved.

\subsection{The effects of parameters $\rm{tan}\beta$, $m_{2}$ and $\mu$ on $R_{D^{(*)}}$}
\begin{figure}[!h]
\centering
\includegraphics[width=8cm]{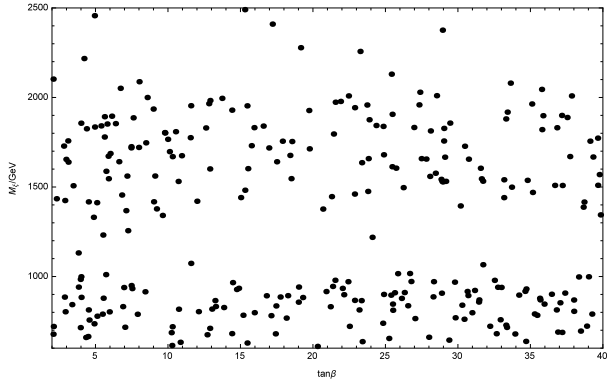}\\
\caption{The allowed parameters in the plane of $M_\xi$ versus $\rm{tan}\beta$ with $g_L=0.45$,\\ $(m_{\tilde{L}}^{2})_{11}=(m_{\tilde{R}}^{2})_{11}=(m_{\tilde{L}}^{2})_{22}=(m_{\tilde{R}}^{2})_{22}=5.5\times10^{5} \rm{GeV^{2}}$, $(m_{\tilde{L}}^{2})_{33}=(m_{\tilde{R}}^{2})_{33}=3\times10^{8} \rm{GeV^{2}}$.
} \label{fig5}
\end{figure}

We also research the effects of parameters $\rm{tan}\beta$, $m_{2}$ and $\mu$ on $R_{D^{(*)}}$.
With the supposition $g_L=0.45$, $(m_{\tilde{L}}^{2})_{11}=(m_{\tilde{R}}^{2})_{11}=(m_{\tilde{L}}^{2})_{22}=(m_{\tilde{R}}^{2})_{22}=5.5\times10^{5} \rm{GeV^{2}}$, $(m_{\tilde{L}}^{2})_{33}=(m_{\tilde{R}}^{2})_{33}=3\times10^{8} \rm{GeV^{2}}$, and $m_{2} = \mu = M_\xi$, we scan the parameters of $M_\xi$ versus $\rm{tan}\beta$ in FIG.~\ref{fig5}.

All the points in  FIG.~\ref{fig5} can make $R_{D}$ ($R_{D^{*}}$) reach 0.304 (0.261), and some particular points can bring $R_{D}$ ($R_{D^{*}}$) to 0.305 (0.262). The theoretical values are improved, but they are not as big as we expected. However, our results are still better than those in the SM.
All of the above discussions only consider the central values in the SM.
If we consider the uncertainty of SM predictions $R_{D}=0.299\pm0.003$ \cite{tianjia1} and $R_{D^*}=0.257\pm0.003$ \cite{tianjia1}, our theoretical value of $R_{D}$ ($R_{D^{*}}$) can reach 0.308 (0.265), when we take the biggest value of the SM prediction.

\section{Summary and future prospects}
The SM cannot well explain the experimental data for $R_{D^{(*)}}$ well, so we hold that SM should be the low energy effective theory of a large model. We think the BLMSSM is more promising for testing in the future. Compared with the MSSM, there are new particles and new parameters in the BLMSSM, and the new contributions from these are the keys to solve the anomalies in $R_{D^{(*)}}$. For instance, the three lepneutralinos $\chi^0_{L}$ are new particles in the BLMSSM, and the Feynman diagram with $\chi^0_{L}$ can give new contributions to $R_{D^{(*)}}$.

We find that the parameters $(m_{\tilde{L}}^{2})_{ii}$ and $(m_{\tilde{R}}^{2})_{ii}$
influence the theoretical results to some extent, and $R_{D^{(*)}e}$ is approximately equal to $R_{D^{(*)}\mu}$  only if there is a certain relationship between parameters $(m_{\tilde{L}}^{2})_{ii}$ and $(m_{\tilde{R}}^{2})_{ii}$. After that, the effect of parameter $g_{L}$ is important, and we can further raise theoretical values when it takes some appropriate values.
Finally, using the central value of the SM prediction we scan the parameter space, and bring the value of $R_{D}$ ($R_{D^{*}}$) to 0.305 (0.262).
Taking into account the SM uncertainty and adopting the biggest value in the SM,  our result for $R_{D}$ ($R_{D^{*}}$) can reach 0.308 (0.265).

In this paper, we use effective field theory to compute $R_{D^{(*)}}$ in the BLMSSM. The one-loop corrections to $R_{D^{(*)}}$ have an effect and the theoretical values can be increased (though they are not big improvements). We notice that the measurements of $R_{D^{*}}$ (see TABLE \ref{tab1}) are not as large as the original measurements. This suggests that $R_{D^{*}}$ perhaps is not so large. From the trend of experimental measurement, the experimental values of $R_{D^{(*)}}$ might be smaller in the future. In fact, without considering this case, the measurement of $R_{D}$ ($R_{D^{*}}$) shows $2.3\sigma$ ($3.1\sigma$)
 deviation from its SM prediction, and our theoretical values are still better than the predictions given by SM.
 On the whole, the problem of $R_{D}$ ($R_{D^{*}}$) should be further researched both experimentally and theoretically in the future.

\section*{Acknowledgement}
Supported by the Major Project of
NNSFC (No. 11535002, No. 11605037, No. 11705045),
the Natural Science Foundation of Hebei province with Grant
No. A2016201010 and No. A2016201069, and the Natural Science Fund of
Hebei University with Grants No. 2011JQ05 and No. 2012-
242, Hebei Key Lab of Optic-Electronic Information and
Materials, the midwest universities comprehensive strength
promotion project.
\appendix
\begin{center}
\Large{{\bf Appendix}}
\end{center}
\vspace{-8mm}
~\\
The formulae for the one-loop integral are:
\begin{eqnarray}
&&\frac{(2 \pi \kappa)^{4-D}}{i \pi^{2}}\int d^{D}p\frac{1}{(p^{2}-m_{1}^{2})(p^{2}-m_{2}^{2})(p^{2}-m_{3}^{2})}=-\frac{1}{\Lambda_{NP}^{2}}F_{11}(x_{1},x_{2},x_{3}),\nonumber\\&&\hspace{0cm}
\frac{(2 \pi \kappa)^{4-D}}{i \pi^{2}}\int d^{D}p\frac{p^{2}}{(p^{2}-m_{1}^{2})(p^{2}-m_{2}^{2})(p^{2}-m_{3}^{2})}=\frac{1}{\varepsilon}-\gamma_{E}+\ln(4\pi \kappa^2/\Lambda^2_{_{NP}})+1\nonumber\\&&\hspace{8.8cm}-F_{21}(x_{1},x_{2},x_{3}),\nonumber\\&&\hspace{0cm}
\frac{(2 \pi \kappa)^{4-D}}{i \pi^{2}}\int d^{D}p\frac{1}{(p^{2}-m_{1}^{2})(p^{2}-m_{2}^{2})(p^{2}-m_{3}^{2})(p^{2}-m_{4}^{2})}=-\frac{1}{\Lambda_{NP}^{4}}F_{11}(x_{1},x_{2},x_{3},x_{4}),\nonumber\\&&\hspace{0cm}
\frac{(2 \pi \kappa)^{4-D}}{i \pi^{2}}\int d^{D}p\frac{p^{2}}{(p^{2}-m_{1}^{2})(p^{2}-m_{2}^{2})(p^{2}-m_{3}^{2})(p^{2}-m_{4}^{2})}=-\frac{1}{\Lambda_{NP}^{2}}F_{21}(x_{1},x_{2},x_{3},x_{4}),\nonumber
\end{eqnarray}
\begin{eqnarray}
&&F_{11}(x_{1},x_{2},x_{3})=\frac{x_{1}\ln x_{1}}{(x_{1}-x_{2})(x_{1}-x_{3})}+\frac{x_{2}\ln x_{2}}{(x_{2}-x_{1})(x_{2}-x_{3})}+\frac{x_{3}\ln x_{3}}{(x_{3}-x_{1})(x_{3}-x_{2})},\nonumber\\&&\hspace{0cm}
F_{21}(x_{1},x_{2},x_{3})=\frac{x_{1}^{2}\ln x_{1}}{(x_{1}-x_{2})(x_{1}-x_{3})}+\frac{x_{2}^{2}\ln x_{2}}{(x_{2}-x_{1})(x_{2}-x_{3})}+\frac{x_{3}^{2}\ln x_{3}}{(x_{3}-x_{1})(x_{3}-x_{2})},\nonumber\\&&\hspace{0cm}
F_{11}(x_{1},x_{2},x_{3},x_{4})=\frac{x_{1}\ln x_{1}}{(x_{1}-x_{2})(x_{1}-x_{3})(x_{1}-x_{4})}+\frac{x_{2}\ln x_{2}}{(x_{2}-x_{1})(x_{2}-x_{3})(x_{2}-x_{4})}\nonumber\\&&\hspace{3.3cm}
+\frac{x_{3}\ln x_{3}}{(x_{3}-x_{1})(x_{3}-x_{2})(x_{3}-x_{4})}+\frac{x_{4}\ln x_{4}}{(x_{4}-x_{1})(x_{4}-x_{2})(x_{4}-x_{3})},\nonumber\\&&\hspace{0cm}
F_{21}(x_{1},x_{2},x_{3},x_{4})=\frac{x_{1}^{2}\ln x_{1}}{(x_{1}-x_{2})(x_{1}-x_{3})(x_{1}-x_{4})}+\frac{x_{2}^{2}\ln x_{2}}{(x_{2}-x_{1})(x_{2}-x_{3})(x_{2}-x_{4})}\nonumber\\&&\hspace{3.3cm}
+\frac{x_{3}^{2}\ln x_{3}}{(x_{3}-x_{1})(x_{3}-x_{2})(x_{3}-x_{4})}+\frac{x_{4}^{2}\ln x_{4}}{(x_{4}-x_{1})(x_{4}-x_{2})(x_{4}-x_{3})},\nonumber
\end{eqnarray}

\end{document}